\documentclass[preprint2]{proto}
\usepackage{natbib}
\bibpunct{(}{)}{;}{a}{,}{,}
\usepackage{times}

\newcommand{\msun}{\mbox{\,M$_\odot$}}

\begin{document}

\title{\textbf{\LARGE The Origin and Universality of the Stellar Initial Mass Function}}

\author {\textbf{\large Stella S. R. Offner}}
\affil{\small\em Yale University}

\author {\textbf{\large Paul C. Clark}}
\affil{\small\em  Universit\"{a}t Heidelberg}

\author {\textbf{\large Patrick Hennebelle}}
\affil{\small\em  AIM-CEA/Saclay}

\author {\textbf{\large Nathan Bastian}}
\affil{\small\em  Liverpool John Moores University}

\author {\textbf{\large Matthew R. Bate}}
\affil{\small\em University of Exeter}

\author {\textbf{\large Philip F. Hopkins}}
\affil{\small\em California Institute of Technology}

\author {\textbf{\large Estelle Moraux}}
\affil{\small\em Institut de Plan\'{e}tologie et d'Astrophysique de Grenoble }

\author {\textbf{\large Anthony P. Whitworth}}
\affil{\small\em Cardiff University}

\begin{abstract}
\baselineskip = 11pt
\leftskip = 0.65in 
\rightskip = 0.65in
\parindent=1pc
{\small We review current theories for the origin of the Stellar Initial Mass Function (IMF) with particular focus on the extent to which the IMF can be considered universal across various environments.  To place the issue in an observational context, we summarize the techniques used to determine the IMF for different stellar populations, the uncertainties affecting the results, and the evidence for systematic departures from universality under extreme circumstances. We next consider theories for the formation of prestellar cores by turbulent fragmentation and the possible impact of various thermal, hydrodynamic and magneto-hydrodynamic instabilities. We address the conversion of prestellar cores into stars and evaluate the roles played by different processes: competitive accretion, dynamical fragmentation, ejection and starvation, filament fragmentation and filamentary accretion flows, disk formation and fragmentation, critical scales imposed by thermodynamics, and magnetic braking. We present explanations for the characteristic shapes of the Present-Day Prestellar Core Mass Function and the IMF and consider what significance can be attached to their apparent similarity. Substantial computational advances have occurred in recent years, and we review the numerical simulations that have been performed to predict the IMF directly and discuss the influence of dynamics, time-dependent phenomena, and initial conditions.
 \\~\\~\\~}
\end{abstract}

\section{\textbf{INTRODUCTION}}

Measuring the Stellar Initial Mass Function (hereafter IMF), and understanding its genesis, is a central issue in the study of star formation. It also has a fundamental bearing on many other areas of astronomy, for example, modeling the microphysics of galactic structure and evolution, and tracking the build-up of the heavy elements that are essential for planet formation and for life. Since it appears that the formation and architecture of a planetary system strongly depends on the mass of the host star, and -- for stars in clusters -- on feedback effects from nearby massive stars, an understanding of the IMF is critical on this count too.

Many theoretical ideas have been proposed to explain the origin of the IMF. Current popular ideas can be roughly divided into two opposing categories: models that are deterministic and models that are stochastic. Simply put, the debate is one of nature (cores) versus nurture (dynamics). 


Theories postulating a direct mapping between the Core Mass Function (CMF) and the IMF are predicated on the idea that stellar masses are inherited from the distribution of natal core masses \citep{PadNor2002,HenCha2008,HenCha2009,Oey2011,Hop2012b}.  
In such models the origin of the IMF is essentially deterministic since they assume that stellar mass is accreted from a local reservoir of gas.  Models for the CMF are based on the ubiquitous presence of energetic, turbulent motions within molecular clouds, which naturally create a distribution of density and velocity fluctuations.  Proponents of such models appeal to the striking similarity between the CMF and IMF shapes.  {\it If} the mapping between the CMF and IMF is sufficiently simple and cores evolve in relative isolation from one another then the problem of predicting the IMF reduces to understanding the CMF.

At the opposite extreme, some models propose that final stellar masses are completely independent of initial core masses, and thus any similarity between the IMF and CMF is coincidental \citep{BonBatClaetal2001,BatBonBro2003,ClaKleBon2007}. Models which emphasize the importance of dynamical interactions and stochastic accretion, for example those based on the idea of stars competing for gas, fall in this category. 
The dynamical view of star formation has its roots in early, simple numerical studies. \cite{Larson1978} was the first to perform numerical simulations of the collapse of a cloud to form a group of protostars and emphasized that  a mass spectrum of objects could be obtained, ``at least in part by accretion processes and the competition between different accreting objects".  Both \cite{Larson1978} and \cite{Zinnecker1982} developed simple analytic models, finding that mass spectra of the form ${\rm d}N/{\rm d} \log M \approx M^{-1}$ (resembling the high-mass end of the IMF) could be produced by protostars competing for the accretion of gas from their parent cloud.  This pioneering work was neglected for close to two decades until more advanced numerical simulations of the formation of groups of protostars could be performed.  

Advances in computational power have subsequently increased the scope and utility of numerical simulations of star formation.  Current state-of-the-art calculations of cluster formation may include a variety of physical effects as well as resolution down to AU scales. A variety of parameter studies have enabled the exploration of the IMF as a function of initial conditions and environment. Although still physically incomplete, these simulations provide tantalizing clues on the origin of the IMF in addition to a wealth of secondary metrics such as gas kinematics, accretion disk properties, stellar velocities, and stellar multiplicities, which provide further constraints on IMF theories.

Meanwhile, challenges persist for the observational determination of the IMF. While observations of resolved stellar populations support a universal stellar IMF, statistics are sparse for both brown dwarfs and very high-mass stars, stellar multiplicity is often unresolved, and masses depend significantly on uncertain stellar evolutionary models.
Recent observations are extending beyond local environments and the Milky Way by employing a variety of techniques.
Persuasive evidence for IMF variations at high-masses has been presented for extreme regions such as the galactic center, giant ellipticals, and dwarf galaxies \citep{VanCon2010,CapMcDAlaetal2012,LuDoGheetal2013,GehBroTumetal2013}. Although statistical and observational uncertainties remain formidable, these observations provide fresh impetus for general and predictive theoretical models.

 The apparent Galactic IMF invariance and the categorical differences between current theories motivates a number of important, open questions that we will discuss in this chapter:  What sets the characteristic stellar mass and why is it apparently invariant in the local universe? Do high-mass stars form differently than low-mass stars and, if so, what implications does this have for their relative numbers? How sensitive is a star's mass to its initial core mass? To what extent do dynamical interactions between forming stars influence the IMF? Finally, what physical effects impact stellar multiplicity and how does this impinge upon star formation efficiency? We will address these questions in the context of current theories and discuss the prognosis for verification and synthesis of the theoretical ideas given observational constraints.

We begin by reviewing the observational determination of the IMF in \S \ref{obs_sec}. In \S \ref{cmf_sec} we discuss the form of the CMF and theories for its origin.  Next, \S \ref{link_sec} considers the possible link between the CMF and IMF. We discuss simulations of forming clusters in \S \ref{cluster_sec} and evaluate the role of dynamics, time-dependent phenomena and initial conditions in producing the IMF. We summarize the status of open questions and present suggestions for future work  in \S \ref{sum_sec}.

\section{\textbf{OBSERVATIONAL DETERMINATION OF THE IMF}}\label{obs_sec}

\subsection{Forms of the IMF}

When interpreting observations and models of the IMF, it is common to compare with standard analytical forms that have been put forward in the literature.  For example, studies focusing on stars more massive than a few $\msun$ commonly adopt a power-law distribution of the form $dN \propto M^{-\alpha}dM$.  This was the form originally adopted by \citet{Salpeter1955}, who determined $\alpha=2.35$ by fitting observational data.  Over 55 years later, this value is still considered the standard for stars above $1\msun$.  However, this function diverges as it approaches zero, so it is clear that there must be a break or turn-over in the IMF at low masses.  This was noted by \citet{MilSca1979} who found that the IMF was well approximated by a log-normal distribution between $0.1$ and $\sim30$\msun, where there is a clear flattening below $1\msun$.  However, current data suggest that a log-normal under-predicts the number of massive stars ($>20$\msun) due to the turn down above this mass, relative to a continuous power-law {\citep[see e.g.][]{BasCovMey2010}.

More modern forms often adopt a log-normal distribution at low masses and a power-law tail above a solar mass \citep{Chabrier2003,Chabrier2005}.  A very similar approach, although analytically different, is to represent the full mass range of the IMF as a series of power-laws, such as in \citet{Kroupa2001,Kroupa2002}.  Above $\sim0.2\msun$, the multiple-power-law segments and log-normal with a power-law tail at high masses agree very well as shown by Figure \ref{imffits}.  However, the form of the IMF at the low mass end is still relatively uncertain and subject to ongoing debate (e.g., \citealt{ThiKro2007,ThiKro2008,KroWeiPfletal2013}).  The philosophical difference between these two forms reduces to whether one believes that star-formation is a continuous process or whether there are distinct physical processes that dominate in certain regimes, such as at the stellar/sub-stellar boundary.

Other proposed functional forms include a truncated exponential where $dN \propto M^{-\alpha}(1-{\rm exp}[(-M/M_p)^{-\beta}]) dM$ \citep{deMPar2001,ParMcKHol2011}, the Pareto-Levy family of stable distributions, which includes the Gaussian \citep{CarWhi2012}, and a log-normal joined with power laws at both low and high masses \citep{Maschberger2013}.  It is usually not possible to discriminate between these forms on the basis of observations due to significant uncertainties.
We note that Figure \ref{imffits} shows only the proposed IMF functions and not the error associated with the derivations of these fits from the data.  In all cases shown, the authors fit the galactic field IMF over a restricted mass range and then extrapolated their functional fit to the whole mass range. Observers and theoreticians now use these functional forms  {\it without any error bars} to evaluate how well their data (or models) are consistent within their uncertainties.


\begin{figure}[t]
\epsscale{1.0}
\vspace{-0.15in}
\plotone{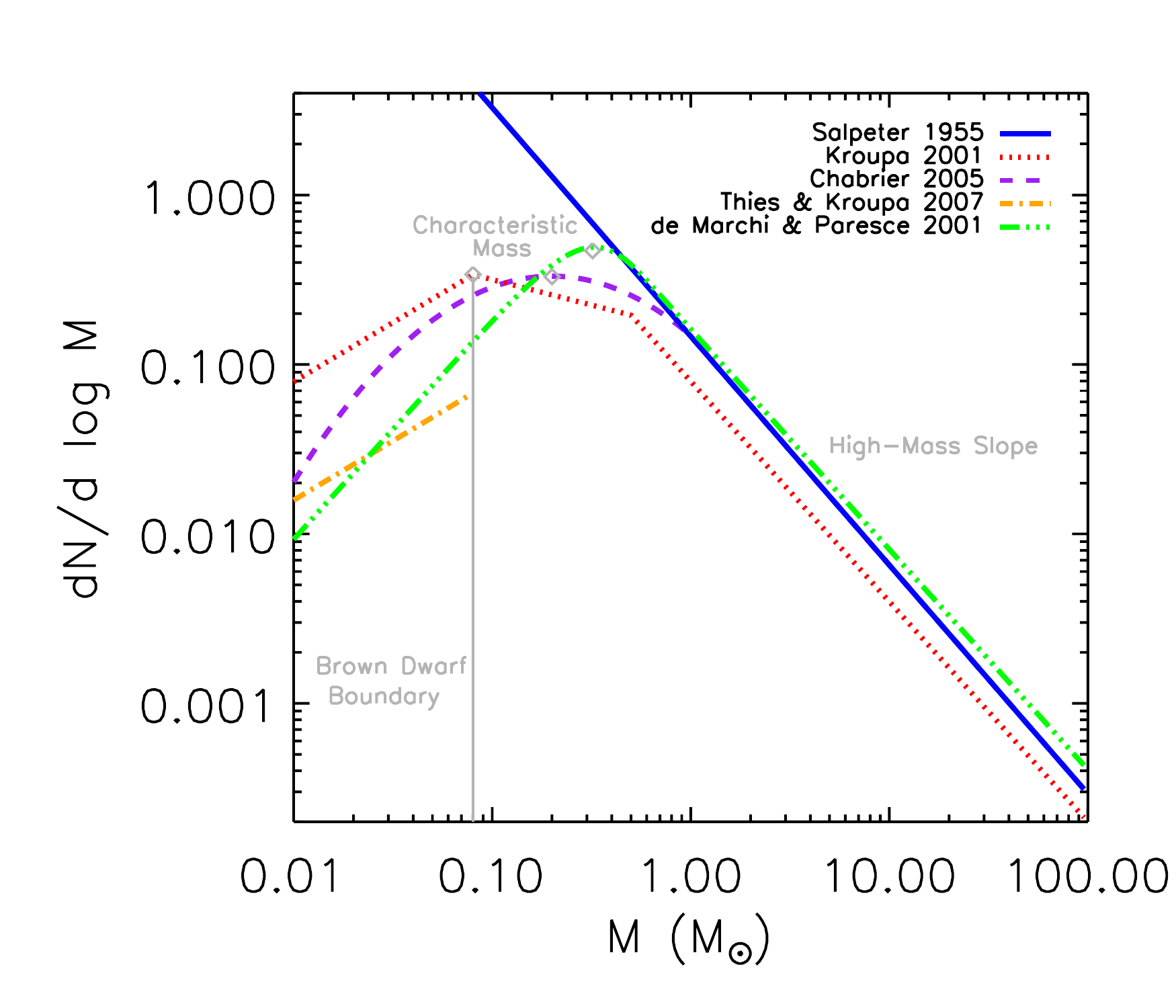}
\vspace{-0.35in}
\caption{IMF functional forms proposed by various authors from fits to Galactic stellar data. With the exception of the Salpeter slope, the curves are normalized such that the integral over mass is unity. When comparing with observational data, the normalization is set by the total number of objects as shown in Figure \ref{imfmw}.
\label{imffits}}
\end{figure}

\subsection{IMF Universality in the Milky Way}

\subsubsection{Uncertainties in the IMF Determination From Resolved Populations}

The basic observational method to derive the IMF consists of three steps. First, observers measure the luminosity function (LF) of a complete sample of stars that lie within some defined volume. Next, the LF is converted into a present day mass function (PDMF) using a mass-magnitude relationship. Finally,  the PDMF is corrected for the star formation history, stellar evolution, galactic structure, cluster dynamical evolution and binarity to obtain the individual-star IMF. None of these steps is straightforward, and many biases and systematic uncertainties may be introduced during the process. A detailed discussion of these limitations at low masses and in the substellar regime is given by \citet{Jeffries2012} and \citet{Luhman2012}.

When obtaining the LF, defining a complete sample of stars for a given volume can be challenging. Studies of field stars from photometric surveys, either wide and relatively shallow \citep[e.g.][]{BocHawCovetal2010} or narrow but deep \citep[e.g.][]{SchRobReyetal2006}, are magnitude limited and need to be corrected for the Malmquist bias. Indeed, when distances are estimated from photometric or spectroscopic parallax, the volume limited sample inferred will be polluted by more distant bright stars due to observational uncertainties and intrinsic dispersion in the color-magnitude relation. This effect can be fairly significant and change the slope of the luminosity function by more than 10\%. Studies of nearby stars, for which the distance is determined from trigonometric parallax, are not affected by this bias, but they are affected by the \citet{LutKel1973} bias, in which the averaged measured parallax is larger than the true parallax, if the parallax uncertainties exceed 10\%. These issues reduce the studies within 20~pc to a completeness level of $\sim$80\% \citep{ReiCruAll2007}, which results in small number statistics, especially at low masses. 
Cluster studies do not have these problems since a complete sample can in principle be obtained from the complete cluster census. However, secure membership cannot be assessed from photometry alone and proper motion measurements as well as spectroscopic follow-up are often necessary. In addition, the remaining contamination by field stars may be large if  cluster properties such as age and proper motion are similar to the field. 

For young star forming regions, differential extinction may be a significant problem. This is usually taken into account by limiting the sample up to a given $A_V$. However, imposing a limit may exclude the more central regions, which are often more extincted, from the determination of the LF. If the stellar spatial distribution is not the same at all masses and, in particular, if mass segregation is present, excluding high $A_V$ sources may introduce a strong bias. Such an effect can be tested by simulating a fake cluster as in \citet{ParMasAlv2012}. 

Even with no extinction, mass segregation must be taken into account because the cluster LF may be different in the center and in the outskirts. The best way to account for this issue is to cover an area larger than the cluster extent. However, in very rich and distant clusters, this may not be sufficient. Photometric surveys are blind in the vicinity of very bright stars since they are not able to detect nearby faint objects, and crowding can result in blended stars. Both effects may produce a strong bias in the LF. All the above considerations clearly show that incompleteness is often a major problem, one that is more severe at low masses, and should be treated carefully when determining the LF.

Once the LF has been obtained, the luminosities must be converted into masses. Different methods may be used,  which convert either absolute magnitude, bolometric luminosities or effective temperature, and which use either empirical or theoretical relationships. For main sequence stars with accurate parallax and photometry, dynamical masses can be measured in  binary systems to give an empirical mass-magnitude relationship \citep[e.g.][]{DelForSegetal2000} that can then be used directly. While there is almost no scatter and there is good agreement with evolutionary models when using near-infrared magnitude, this is clearly not the case in the V-band due to opacity uncertainties and metallicity dependence, especially at lower masses. 

Another issue to consider is the age dependence of the mass-luminosity relation and the difficulty of estimating stellar ages unless the stars are coeval. The population of field stars has a large age spread, and many of the oldest stars have already left the main sequence. In this blended population, the simplest assumption is that all field stars are approximately the age of the universe. In this case, stars more massive than 0.8$\msun$ have already left the main sequence while younger low-mass stars ($M<0.1\msun$), which take a considerable amount of time to contract, are still on the pre-main sequence e.g., \citealt{BarChaAlletal1998}).
In the substellar domain, the situation is more complicated because brown dwarfs never reach the main sequence. They never initiate hydrogen fusion but simply cool with time, which leads to a degeneracy between mass, effective temperature and age. In principle, position in the HR diagram could determine both mass and age, but different evolutionary models lead to large differences in the deduced masses. The discrepancy between the \citet{BurMarHubetal1997} and \citet{ChaBarAlletal2000} models is mainly around 10\% but can be as large as 100\% \citep{KonGheBaretal2010}. Uncertainties in the models may be due to the formation/disappearance of atmospheric clouds, equation of state, assumptions for the initial conditions and accretion \citep{HosOffKru2011,BarVorCha2012}, treatment of convection, and effect of magnetic fields \citep{ChaKuk2006}. We also note that even if a model is very accurate, a typical measurement uncertainty of 100~K in the effective temperature may correspond to a factor of two difference in mass \citep[see e.g.][]{BurMarHubetal1997}.

The last step in the process, which consists of deriving the IMF from the PDMF, is also subject to many biases as it often relies on various models and assumptions for the star formation history, Galactic structure, and stellar evolution. For example, short-lived massive stars are under represented in the field population since they evolve to become white dwarfs or neutron stars. Consequently, an estimate of the star formation history is required to transform the field PDMF to the IMF above 0.8$\msun$. Moreover, the low mass stars are older on average and are more widely distributed above and below the Galactic plane. This must be corrected for, especially for nearby surveys that do not explore the full Galactic disk scale height. In contrast to the field population, clusters contain stars of similar age, distance and initial composition. This is advantageous since we expect the PDMF of young clusters to be the IMF. However, below 10 Myr, reliable masses are difficult to obtain due to high and variable extinction, reddening, uncertain theoretical models \citep[e.g.][]{BarChaGal2009,HosOffKru2011} and possible age spread. At older ages the PDMF is more robust since it is less affected by these uncertainties, however, it may no longer resemble the IMF. 

As discussed above, mass segregation introduces a radial dependence to the MF that must be taken into account if the survey does not include the entire cluster. This is true for all clusters of any age. Indeed many young clusters appear mass segregated, whereby the most massive stars are more concentrated toward the center \citep[e.g.,][]{HilHar1998}, indicating that mass segregation may be primordial or can occur dynamically on a very short timescale \citep[e.g.,][]{AllGooParetal2010}. More importantly, secular evolution due to 2-body encounters leads to a preferential evaporation of lower mass members, which shifts the peak of the MF towards higher masses \citep{DeMParPor2010}. However, $N$-body numerical simulations suggest this effect is not important when the cluster is younger than its relaxation time: $t_{rlx}= N R_h / 8 \ln N \sigma_V$ \citep[e.g.,][]{AdaDavJametal2002}, where $N$ is the number of star systems in the cluster, $R_h$ is the half mass radius, and $\sigma_V$ is the velocity dispersion.

Finally, another important limitation is that binaries are not resolved in most of these studies. Consequently, the cluster MFs as well as the field MF obtained from deep surveys correspond to the {\it system} MF not the individual-star MF. Correction for binarity can be done for the field stellar MF but not for the substellar domain since the binary properties of field brown dwarfs are fairly uncertain. There is usually no attempt to correct for binarity in clusters, mainly to avoid introducing additional bias since binary properties may depend on age and environment.

\subsubsection{The Field IMF}

Since the seminal study of \citet{Salpeter1955}, the Galactic field stellar IMF is usually described as a power-law with $\alpha=2.35$ above $1 \msun$, although it is  occasionally claimed to be steeper at higher masses \citep[see e.g.][]{Scalo1986}. Between 0.1 and 0.8$\msun$, the PDMF corresponds to the IMF and is now relatively well constrained thanks to recent studies based on local stars with Hipparcos parallax \citep{ReiGizHaw2002} and the much larger sample of field stars with less accurate distances \citep{BocHawCovetal2010,CovHawBocetal2008,DeaNelHam2008}. The results of these studies are in reasonable agreement for $M<0.6\msun$ and suggest the single star IMF at low masses is well described by a power-law with  $\alpha\sim1.1$.  The IMF obtained by \citet{BocHawCovetal2010} suggests a deficit of stars with $M>0.6\msun$ and is better fitted by either a log-normal with a peak mass around $0.18 \msun$ and $\sigma=0.34$ or a broken power law. However, the evidence for a peak is weak on the basis of this mass range alone and the perceived deficit may be due to neglecting binary companions in systems with total mass above $0.8\msun$ \citep{ParMcKHol2011}. Although the statistical uncertainties are much smaller for extended samples than for small local samples of field stars, the systematic errors are larger, yielding similar final uncertainties in both cases. Overall, most results are converging on a field IMF in the stellar domain that has a Salpeter-like slope above $\sim 1 \msun$ and flattens significantly below. This distribution can be equally well represented by segmented power-laws \citep{Kroupa2001} or by a log-normal function with peak mass of $0.15-0.25 \msun$ \citep{Chabrier2005}. 

In the substellar domain the shape of the Galactic disk 
IMF is more uncertain. Instead of converting a LF directly to a MF, which is hampered by the brown dwarf age-dependence in the mass-luminosity relationship, the IMF is usually estimated by predicting the LF via Monte-Carlo simulations and assuming an IMF functional form and a star formation history  \citep[e.g.][]{Burgasser2004}. Recent results from WISE \citep{KirGelCusetal2012} and UKIDSS \citep{BurCarSmietal2013,DayMarPinetal2013} using this approach indicate that the substellar IMF extends well below $10\, {\rm M}_{Jup}$ and may be represented by a power law with $-1<\alpha<0$, although it is still consistent with a log-normal shape due to the large uncertainties. 

\subsubsection{The IMF in Young Clusters}

Several studies have been carried out in young open clusters with ages of 30-150 Myr to determine the IMF across the substellar boundary ($0.03\, \msun \lesssim M \lesssim 0.3\,\msun$).  Since the comprehensive review by \citet{Jeffries2012}, IMF determinations have been obtained using the UKIDSS Galactic Cluster Survey, which includes IC4665 \citep{LodDeWCaretal2011}, the Pleiades  \citep{LodDeaHam2012a}, and  $\alpha$~Per \citep{LodDeaHametal2012b}. The results agree well with previous studies, indicating that the {\it system} IMF is well represented
 by a power-law with $\alpha\simeq 0.6$ in the mass range $0.03-0.5 \msun$ or a log-normal with a peak mass around $0.2-0.3 \msun$ and $\sigma\simeq0.5$, consistent with \citet{Chabrier2005}. Results obtained in star forming regions \citetext{ONC, $\rho$-Oph, NGC2024, NGC1333, IC348, $\sigma$-Ori, Cha I, NGC6611, $\lambda$-Ori, Upper Sco; see \citealt{BasCovMey2010} and \citealt{Jeffries2012} for references as well as \citealt{Lodieu2013,AlvMorBouetal2013,AlvMorBouetal2012,PenBejZapetal2012,SchMuzGeeetal2012,MuzSchGeeetal2012,MuzSchGeeetal2011,GeeSchJayetal2011,BayBarStaetal2011} for more recent studies} are also in remarkable agreement in the same mass range (see Figure \ref{imfmw}). 
 The only clear exception is Taurus, which shows an excess of $0.6-0.8 \msun$ stars and peaks at higher masses \citep{LuhMamAlletal2009}. However, note that an agreement between the {\it system} IMFs does not necessarily imply agreement between the {\it individual} star IMFs, since binary properties may be different.

 At lower masses, the log-normal form may remain a good parametrization, although this is currently debated, since an excess of low-mass brown dwarfs has been reported in Upper Sco \citep{Lodieu2013} 
 and in $\sigma$-Ori \citep{PenBejZapetal2012}, see Figure \ref{imfmw}. 
 These excesses may result from uncertainties in the mass-luminosity relation at very low masses and young ages. 
 
Obtaining measurements of the IMF at high stellar masses ($M \simeq 3-100 \msun$)  is also challenging.
Recent studies since the review from \citet{BasCovMey2010} include the young, massive clusters NGC~2264 \citep{SunBes2010}, Tr~14 and
Tr~16 \citep{HurSunBes2012}, and NGC~6231 \citep{SunSanBes2013}. These
regions exhibit a high-mass slope that is consistent with Salpeter, although the IMF is found to be slightly steeper for NGC~2264
where $\alpha=2.7\pm0.1$. 
In contrast, the MF appears to vary spatially in the much denser and more massive NGC~3603 and Westerlund~1 clusters, as well as in clusters within the Galactic center
such as the Arches, Quintuplet, and young nuclear star clusters.
The high-mass slopes of these extreme regions appear to be consistently flatter than the standard IMF in the inner part of the cluster
and steepen toward the outskirts. This is a strong indication that
these clusters are mass segregated despite their youth. While the global
IMF 
of the Arches is found to be consistent with a
Salpeter IMF \citep{HabStoBraetal2012}, the IMF of Westerlund~1 is controversial since 
\citet{LimChuSunetal2013} report a flatter IMF while \citet{GenBraStoetal2011} report a normal IMF.  In the Quintuplet cluster, \citet{HusStoBraetal2012} derived an IMF that is flatter 
than Salpeter,  but they only investigated the central part of the cluster. In
the young nuclear cluster in the Galactic center,
\citet{LuDoGheetal2013} found a MF slope of $\alpha = 1.7 \pm 0.2$,
which is again flatter than Salpeter. They claim that this region is one
of the few environments where significant deviations have been found
from an otherwise near-universal IMF.

\begin{figure}[t]
\epsscale{1.0}
\vspace{-0.15in}
\plotone{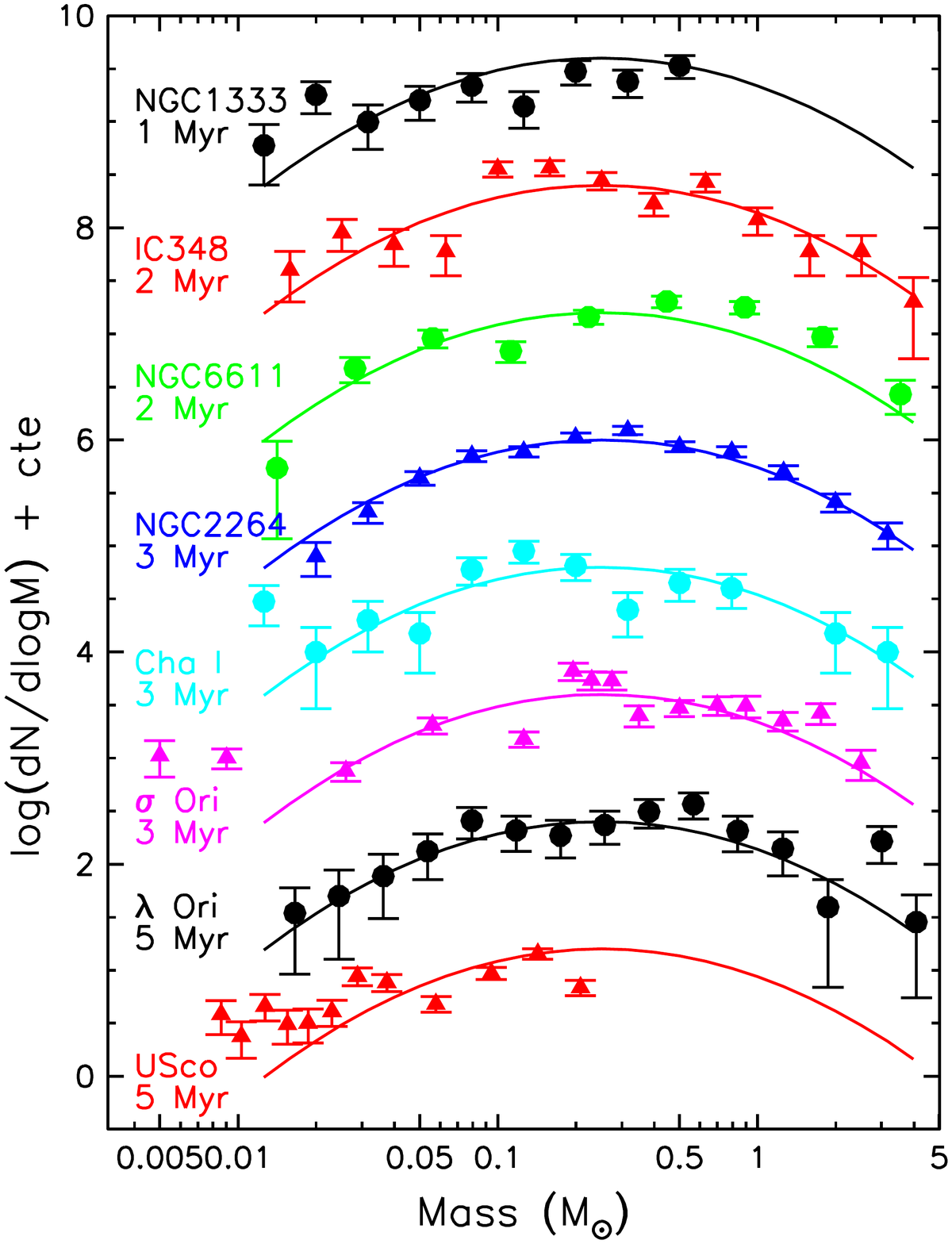} 
\vspace{-0.3in}
\caption{Recent IMF estimates for 8 star forming regions: NGC1333,
\citealt{SchMuzGeeetal2012}; IC348, \citealt{AlvMorBouetal2013};
NGC6611, \citealt{OliJefVan2009}; NGC2264, \citealt{SunBes2010}; Cha~I,
\citealt{Luhman2007}; $\sigma$ Ori, \citealt{PenBejZapetal2012}; $\lambda$
Ori, \citealt{BayBarStaetal2011} and Upper Sco, \citealt{Lodieu2013}.
The error bars represent the Poisson error for each data point. The
solid lines are {\it not} a fit to the data but the log-normal
form proposed by \citet{Chabrier2005} for the IMF, normalized to best
follow the data.
\label{imfmw}}
\end{figure}

\subsection{IMF Universality in the Local Universe}

With high-resolution imaging provided by the Hubble Space Telescope (HST), we can resolve individual stars out to a few Mpc.  This technique has proven to be extremely powerful for determining the star-formation histories of galaxies, and in some cases, the stellar IMF can be directly constrained.  In general, these studies are limited to the high mass portion of the IMF ($M>1$\msun)  simply due to detection limits, however, a few recent studies have attempted to push significantly past  this limit.  It is beyond the scope of this review to discuss the plethora of studies that have measured the PDMF/IMF through resolved stellar populations. Instead, we highlight just a few recent studies and refer the reader to \citet{BasCovMey2010} for a more thorough compilation. See the IMF review by \citet{KroWeiPfletal2013} for an alternative discussion of observational variations.

Increasing distance amplifies the potential caveats and biases present in studies of local regions, e.g., crowding, photometric errors, unresolved binaries, and mass segregation.  \citet{WeiFouHogetal2013} have supplied an overview of these plus additional biases and potential pitfalls and have provided the statistical tools to account for them.  These authors have collated observations of 89 young clusters and star forming regions from the literature and have re-calculated the index of the MF (assuming that it is a power-law) given the statistical and observational uncertainties.  They find that the best-fit  $\alpha$ of 2.46 with a $1\sigma$ dispersion of 0.34, which is similar to Salpeter. Additionally, the authors warn that deriving a physical upper stellar limit to the IMF is difficult and fraught with biases and, consequently, could not be determined from their collected sample.

HST/WFC3 observations of the young, $\sim10^5\,\msun$ cluster Westerlund 1 
have measured its MF down to $\sim0.1\,\msun$.  While mass segregation is certainly present and the inner regions are somewhat incomplete, the overall MF is consistent with the expectation of \citet{Chabrier2005}, i.e., a peak around 0.2-0.3$\msun$ and a power-law at higher masses with a nearly Salpeter slope ({\it Andersen et al.~}in prep). 

Using extremely deep HST images, a few recent studies have attempted to exceed the $1\,\msun$ boundary in extragalactic studies.  For example, \citet{KalAndDotetal2013} have studied a single HST field in the SMC and have been able to constrain the MF from $0.37\,\msun$ to $0.93\,\msun$.  They find that the data are well represented by a power-law with index of 1.9.  This is near the index implied by a \citet{Chabrier2005} IMF for an upper mass of $0.8-0.9\,\msun$.  Surprisingly, a single index for the IMF was able to reproduce their observations over this mass range, i.e., they did not find evidence for a flattening at lower masses as expected for a \citet{Chabrier2005} IMF.  For example, at $0.5\,\msun$, the expected slope is $1.5$ which is within the observational uncertainties.  Hence, the results presented in \citet{KalAndDotetal2013} only differ from the \citet{Chabrier2005} expectation below $\sim0.45\,\msun$ and then by only $1-2\sigma$.  This highlights the difficulty of fitting a single power-law index to the data when a smooth distribution with a continuously changing slope is expected.

We conclude that currently there is no strong evidence (more than $2\sigma$) for IMF variations within local galaxies as determined from studies of resolved stellar populations.


\subsection{Extragalactic Determinations of the IMF}

While techniques to infer the IMF in extragalactic environments are necessarily more indirect than those used locally, the ability to sample more extreme environments provides a promising avenue to discover systematic variations.  We refer the reader to \citet{BasCovMey2010} for a summary of extragalactic IMF studies 
that were published before 2010.  These authors concluded that the majority of observations were consistent with a normal IMF and that many of the claims of systematic IMF variations were potentially due to the assumptions and/or complications, such as extinction corrections and systematic offsets in star-formation rate indicators, in the methods used.  

Since 2010, there has been a flurry of IMF studies focusing on ancient early type (elliptical) galaxies.  These studies have employed two main techniques. Either they dynamically determine the mass of galaxies in order to compare their mass-to-light ratios to the expectations of stellar population synthesis (SPS) models (e.g., \citealt{CapMcDAlaetal2012}) or they use gravity sensitive integrated spectral features to determine the relative numbers of low-mass ($M<0.5\msun$) stars (e.g., \citealt{VanCon2010}).  Contrary to previous studies that found an over-abundance of high mass stars, i.e., a ``top-heavy" IMF, in the high-redshift progenitors of these systems \citep[e.g.,][]{Dave2008}, the above studies have suggested an over-abundance of low-mass stars, i.e., a ``bottom-heavy" IMF, which is consistent with a Salpeter slope (or steeper) down to the hydrogen-burning limit.   
However, it is important to note that while both types of studies find systematic variations, namely that galaxies with higher velocity dispersion have more bottom-heavy IMFs, the reported variations are much smaller than had been previously suggested. The most recent results are within a factor of $\sim2-3$ in total mass for a given amount of light.

One potential caveat is that the above analyses depend on comparing observed properties with expectations of SPS models.  While a large amount of work has gone into developing such models, i.e., by investigating stellar evolution and atmospheres,  it is necessary to calibrate the models with known stellar populations.  For old populations, this is generally done with globular clusters (e.g., \citealt{MarGreRen2003,ConGun2010}). However, there is a deficit of resolved globular clusters in the Galaxy and Magellanic Clouds  
that reach the relatively high metallicity ($Z\ge {1}/{2}\,{\rm Z}_{\odot}$) of elliptical galaxies.  Hence, it is necessary to extrapolate the SPS models outside the calibrated regime.  Using the higher metallicity globular cluster population of M31,  \citet{StrCalSet2011} tested the validity of SPS extrapolations 
and found systematic offsets between the observations and SPS model predictions.  Interestingly, the offset goes in the {\em opposite} direction to that inferred from the dynamical studies of elliptical galaxies, which may imply larger IMF variations.

We conclude that while elliptical galaxies show intriguing evidence for systematic IMF variations, which are potentially due to their unique formation environment, uncertainties underlying the current techniques require further study to confirm these potential systematic variations.

\section{\textbf{THE FORM AND ORIGIN OF THE CMF}}\label{cmf_sec}


Observations of the IMF 
necessarily
occur after the natal gas has formed stars or been dispersed. However,
examining the properties of molecular clouds and the star formation within
provides clues for the initial conditions of star formation and how gas
collapses and produces the IMF.

Within molecular clouds, observers find dense
condensations, often containing embedded protostars,
which appear to have masses distributed like the IMF. A natural
possibility is that these ``cores'' serve as gas reservoirs from which stars accrete most of their mass. If stellar masses are indeed largely determined by the amount of gas in their parent cores then understanding the origin and distribution of core masses becomes a fundamental part of understanding the IMF.

Before continuing we define several important terms that we use throughout the chapter.
Henceforth, a {\it core} refers to a blob of gas which will form one to a few stars, i.e.,  a single star system.  A {\it prestellar core} is one that is expected to undergo collapse but is not yet observed to contain a protostar.  In contrast, a {\it clump} refers to a larger collection of gas, which will likely form a cluster of stars. A {\it star forming clump} specifically indicates 
a clump that is observed to contain several forming stars, while {\it CO clumps} are simply large interstellar
clouds that are observed in CO.  The exact meaning of these terms varies widely in the literature, and we discuss the observational challenges of core and clump definition further in \S\ref{link_sec}.

The possible importance of cores in the star formation process 
leads to a central question:
What determines the mass reservoir that is eventually accreted by a star and when is 
it determined (before the star forms or concomitant with it)? A natural possibility is that 
this reservoir is broadly traced by the population of observed, dense prestellar cores. 
In this scenario, observationally identified cores 
constitute a large part of the final stellar mass. 
This does not preclude gas expulsion by winds, division of mass 
into binary and multiple systems, and further core accretion from the surrounding medium during collapse. These processes are all expected to have some influence but are assumed to be secondary in importance to the core masses.  We dedicate this section to discussing observations and theories for the CMF.

\subsection{Observational Determination of the CMF}

The first determination of the CMF  was achieved 15 years ago  
 using millimeter continuum dust emission in nearby molecular clouds, namely  
$\rho$-Oph \citep{MotAndNer1998,JonWilMoretal2000}
and Serpens \citep{testisargent1998}.
In these studies, cores were identified from the emission data using either a wavelet analysis \citep{LanWilAnd1993,StaMurBij1995} or ``ClumpFind" \citep{WildeGBli1994}, a procedure which separates objects along minimum flux boundaries.   
About 30-50 objects
were identified in each case with masses spanning a wide range ($\sim 0.1\, \msun$ - $10\,\msun$).
These condensations appear sufficiently dense and compact to be gravitationally 
bound. 
The inferred core radial densities suggest that they are similar to Bonnor-Ebert spheres with a constant density in the central region. 
All three studies found a similar core mass spectrum.
Below $0.5\,\msun$, the mass distribution scales as $dN/dM \propto M^{-1.5}$, while above 
$\sim0.5\,\msun$, the distribution is significantly steeper and consistent 
with a power-law  $dN/dm \propto M^{-\alpha}$ where $\alpha \simeq 2-2.5$.
The authors cautioned that the low-mass part of the distribution was likely  
affected by insufficient sensitivity and resolution. 
The inferred high-mass distribution is significantly steeper than the 
mass spectrum inferred for CO clumps which have $\alpha \simeq 1.7$
\citep{Heithausenetal1998,Krameretal1998}. The authors noted that 
the steep mass spectrum was reminiscent of the Salpeter IMF slope 
and proposed that the observed cores were indeed the reservoirs that would contribute most of the eventual stellar masses. More recent observations
by  \citet{Enochetal2007} found similar results for the CMFs 
of Perseus and 
Ophiuchus but a shallower mass distribution for Serpens ($\alpha \simeq 1.6$). 
 
The CMF has also been inferred in the Pipe Nebula 
\citep{AlvLomLad2007,Roman-Zunigaetal2010}
using near-infrared extinction maps
produced from the 2MASS point source catalogs \citep{Lombardietal2006}. 
This technique
nicely complements the results based on dust emission because it is 
independent of dust temperature. The cores identified in the Pipe Nebula
are all starless and much less dense than the cores previously observed in emission, suggesting 
that they may be at a very early evolutionary stage. For masses larger than about $3\,\msun$, the mass spectrum 
is again very similar to the Salpeter IMF, but at lower masses it displays a flat profile with 
a possible peak around $1\,\msun$. Altogether the mass distribution has a shape similar to the 
Chabrier and Kroupa IMFs but is shifted towards larger masses by a factor of $\sim3$.
\citet{NutWar2007} obtained similar results in Orion.  The CMF offset from the IMF has often been 
interpreted as a core-to-star conversion efficiency of about ${1}/{3}$.

The most recent determination of the CMF has been obtained  for the Aquila rift cloud complex using {\em HERSCHEL} data \citep{AndMenBonetal2010,Konyvesetal2010}. 
Since more than $500$ cores have been detected, the statistics 
are about one order of magnitude better than in the previous studies. Most of the cores appear to be  gravitationally bound, meaning that the estimated core masses are comparable to or larger than the local Jeans mass.
 The {\em HERSCHEL} data confirm the conclusions of the previous studies: 
for high masses the CMF slope is comparable to Salpeter, while at low masses the CMF turns over
with a peak located at $\sim1\,\msun$. 
The CMF has also been inferred in the Polaris cloud, which 
is much less dense than Aquila and is non-star forming. 
Interestingly, the identified cores do not appear to be self-gravitating, but the core mass spectrum 
still resembles a log-normal, but it peaks at a much smaller 
value ($\simeq 2 \times 10^{-2}\, \msun$, \citealt{AndMenBonetal2010}).
In summary, the observed CMF in star-forming regions can 
generally be fitted with a log-normal and  resembles the IMF but shifted to higher masses by a factor of $3$.

Finally, we note that cores within Infrared Dark Clouds (IRDCs) also have a mass distribution that is similar to the IMF shape. 
\citet{PerettoFuller2010} derived a clump mass spectrum from the Spitzer 
GLIMPSE data catalog 
of IRDCs \citep{PerettoFuller2009}. 
They found that while the mass spectrum of the IRDCs themselves 
was similar to the mass distribution inferred 
for CO clumps ($\alpha \simeq 1.7$), the 
mass spectrum of the embedded structures was 
steeper and comparable to the Salpeter IMF (although see also \citealt{RagBerGut2009} and \citealt{WilWarKiretal2012}).
Unfortunately,  their observational completeness limit did not allow a determination of the distribution
peak.  The discrepant slopes of core and clump distributions may indicate a transition scale, which we discuss below in the context of theoretical models.

Fifteen years of investigation and increasingly high quality data have reinforced 
the similarity between the shapes of the CMF and the IMF
at large masses. At lower masses, 
a change in the slope of the mass spectrum around a few solar masses appears to be a common feature of many studies. 
However, a definitive demonstration of such a peak will require 
better observational completeness limits.

\subsection{Analytical Theories for the CMF}

If the distribution of stellar
masses is in some way inherited from the distribution of core masses then
theories predicting the CMF are an important first step in understanding
the IMF.  At this point, we stress that the theories presented below 
attempt to infer the mass of the gas reservoir that eventually ends up in the stars rather 
than the masses of cores as defined observationally. Indeed these latter are transient objects, whose masses are time-dependent, which are difficult to properly model 
using a statistical approach. 
In some circumstances, such as in especially compact and clustered regions, discrete cores are difficult to identify observationally. However, individual ``reservoirs'' (collapsing regions of gas from which an embedded protostar will accrete most of its mass) can still be well-defined theoretically.
In the following section, for convenience we use CMF to refer to the mass distribution 
of the stellar mass reservoirs even though these objects are likely somewhat different than observed cores.

Turbulence is ubiquitous in the interstellar medium and appears to be intimately bound to the process of star formation \citep{MacKle2004,ElmSca2004,ScaElm2004}. Most theories of the CMF are predicated on the dual effects of turbulence. Supersonic turbulent motions
can both disperse gas and provide support against gravity \citep{BonHeyFaletal1987}. In the ISM, turbulent support appears to be extremely important, since turbulent motions approximately balance gravity within giant molecular clouds (GMCs) and other cold, dense regions where the thermal pressure is negligible \citep{Eva1999}. However, 
this support cannot simply be treated as hydrostatic ``pressure.'' The same turbulent motions necessarily generate compressions, shocks, and rarefactions; since these effects are multiplicative, the central limit theorem naturally leads to a log-normal density distribution whose width increases with the Mach number, $\mathcal{M}$ (see \citealt{Vaz1994}, \citealt{PadNorJon1997}; although some non-log-normal corrections are discussed in \citealt{Hop2013b}). Because density fluctuations grow exponentially while velocities grow linearly, the presence of supersonic turbulence will always produce dense, {\em local} regions that are {\em less} stable against self-gravity, even if turbulent motions globally balance gravity.

\citet{PadNorJon1997} presented one of the first models to combine the effects of density fluctuations and gravity to predict a CMF. Assuming a log-normal density probability density function (PDF) and isothermal gas, they estimated the number of regions in which the density would be sufficiently high so that gravity overwhelmed thermal support. They further assumed these regions should be analogous to observed cores. Their method naturally predicts a log-normal-like CMF with a peak and a power-law extension at high masses. 
In this calculation, the peak of the IMF comes from the log-normal density PDF, which
stiffly drops at high density. The slope of the distribution at high masses reflects
the counting of thermal Jeans masses in a medium that follows a log-normal density distribution.
The result is similar to observed CMFs, although both the high and low-mass slopes predicted appear to be shallower than those observed. \citet{PadNorJon1997} noted two main limitations to their derivation. First, they neglected local turbulent support. Second, gravitational instability, and the actual mass of a region, depends not just on the density and temperature but also on the size scale. \citet{PadNor2002} extended this method by adopting the simple heuristic assumption that the size of each region followed the characteristic size scale of post-shock gas, which could be computed given some $\mathcal{M}$. They argued that this more naturally led to a power-law high-mass slope.

\citet{Inutsuka2001} was the first to approach the problem of the CMF with the formalism from \citet{PreSch1974}, which is used within cosmology to calculate the mass function of dark matter halos. 
\citet{HenCha2008,HenCha2009} then extended this previous work in more detail.  As in \citet{PreSch1974}, \citet{HenCha2008} assumed the gas density is a Gaussian random field, although in the context of turbulent clouds the field becomes log-density rather than linear-density. If the 
density power spectrum is known, the density can be evaluated for all scales, resolving the second limitation above. The 
fluctuations on all scales can be compared to some threshold to assess gravitational boundedness and, thus, calculate the mass spectrum of self-gravitating turbulent fluctuations above some simple over-density.
 By considering the importance of turbulent as well as thermal support in defining self-gravitating regions,  \citet{HenCha2008} predicted a CMF with a peak, a Salpeter-like power-law at high masses, and a rapidly declining (``stiff'') distribution at low masses. 
This multi-scale approach provides one solution for the ``cloud-in-cloud" problem identified by \citet{BonColEfsetal1991}.
This problem arises because small scale structures are generally embedded within larger ones, which complicates structure counting. The original
formulation by \citet{PreSch1974} suffered from this
problem and led them to overestimate the number of structures by a factor of 2 \citep{BonColEfsetal1991}.
The \citet{HenCha2008} approach also accounts for turbulent fragmentation but not for gravitationally induced fragmentation, which is much more difficult to include.
\citet{HenCha2008} found that their prediction for non-self gravitating masses agreed well with the observed mass spectrum of CO clumps but was less consistent with the IMF-like observed CMF.

\citet{Hop2012a,Hop2012b} extended this approach even further to encompass the entire galactic disk and include rotational support. This invokes ``excursion set theory,'' which is essentially the theory of the statistics of random walks. Excursion set theory was first applied to cosmology by \citet{BonColEfsetal1991} to make the Press \&\ Schechter approach mathematically rigorous and solve the cloud-in-cloud problem (for a review of its current applications in cosmology, see \citealt{Zen2007}). In short, imagine constructing a statistical realization of the density field. In the context of star formation, one can use the density power spectrum of supersonic turbulent gas to correctly include contributions from flows on different scales. Hopkins showed that the density variance versus scale, which is the most important quantity needed to construct the density field, naturally follows from the largest Mach numbers in the galactic disk. Next, one picks a random 
point in the field and smooths the gas density around that point in a ``window'' of some arbitrary radius. As this radius increases from zero to some arbitrarily large value, the mean density in the window can be compared to some threshold governing the self-gravitating collapse of the gas. The threshold may be computed by including thermal, turbulent, magnetic, and rotational support and varies as a function of the size scale. If the field crosses this threshold at any point, then there is both a minimum and a maximum ``window radius'' in which the smoothed field is above the threshold. Within this window, the field may cross back-and-forth multiple times. 
Excursion set theory naturally resolves the cloud-in-cloud problem by explicitly predicting sub-structures inside of larger structures. And indeed, because of the very large dynamic range of turbulence in galaxies, self-gravitating objects inherently include self-gravitating sub-structures on a wide range of scales. 

Hopkins defined the ``first crossing''  as the largest ``smoothing scale'' for which a structure is self-gravitating and demonstrated that the predicted mass spectrum and properties such as the size-mass and linewidth-size relations of the objects agreed very well with those of molecular clouds. At the other extreme, on small scales where thermal support becomes important, there is always a minimum self-gravitating scale, the ``last-crossing'', within a structure. \citet{Hop2012a} showed that the distribution of structures defined by the last-crossing agreed well with the CMF. In fact, under the right circumstances, the predicted last-crossing distribution reduces to the Hennebelle \&\ Chabrier formulation.  Figure~\ref{hopkins} illustrates the predictions of the various
theories and shows the corresponding mass functions. Because the turbulent cascade is set by the global galactic disk properties, this approach naturally predicts an invariant CMF/IMF at high masses within the Milky Way \citep{Hop2013d}. \citet{Hop2013c} also noted that this formalism also 
predicts the clustering of cores, which is analogous to galaxy clustering. Consequently, these statistics provide a quantitative explanation for the large-scale clustering in star formation. 

Statistical tools, such as excursion set theory, are the theoretical analog of new observational metrics such as ``dendrograms'';  both seek to describe the statistical hierarchy of self-gravitating structures as a function of scale \citep{RosPinKauetal2008,GooRosBoretal2009}. Comparing theory with observations requires the additional step of smoothing the observed gas on various scales and accounting for projection effects, which may be challenging (e.g., \citealt{BeaOffSheetal2013}). However, by constructing the statistics of self-gravitating objects defined on their smallest/largest self-gravitating scales, it is possible to rigorously define the observational analog of last/first crossings. One can imagine even more powerful constraints can be obtained by using these methods to directly compare the behavior of the density and velocity fields as a function of smoothing scale (e.g., by examining power spectra, covariance, the spatial correlation function of self-gravitating regions on different scales, or rate of ``back-and-forth'' crossings as a function of scale). This would allow observations and theory to compare the entire dynamic range of turbulent fragmentation {\it using all the information in the field} rather than focusing on some, ultimately arbitrary, definition of a ``core'' and a ``cloud.''

In summary, these theories provide a natural link between 
the CMF,  
GMC mass function, and star cluster mass function for a wide range of systems. This link derives from the inherently scale free physics of turbulent fluctuations plus gravity. The most robust features of these models are the Salpeter-like slope ($\alpha\sim2.2-2.4$),
which emerges generically from scale-free processes, and the log-normal-like turnover, which emerges from the central limit theorem. The location of the turnover, in any model where turbulent density fluctuations are important, is closely tied to the sonic length, which is the scale, $R_{s}$, below which thermal or magnetic support dominates over turbulence. Below this scale density fluctuations become rapidly damped and cannot produce new cores. As a result, the location of the peak, and the rapidity of the CMF turnover below it, depend dramatically on the gas thermodynamics \citep{Larson2005,JapKleLaretal2005}.
Unlike what is often assumed, however, the dependence on thermodynamics displayed by turbulent models and simulations with driven turbulence is not the same as a Jeans-mass dependence (see \citealt{HenCha2009}).  The difference between the scaling of the Jeans mass, $M_J \simeq c_{s}^{3}/(\rho^{1/2}\,G^{3/2})$,  and the sonic mass, $M_s \simeq v_{\rm turb}^{2}\,R_{s}/G$, can be dramatic when extrapolating to extreme regions such as starbursts where the turbulent velocity, $v_{\rm turb}$, is large  \citep{Hop2013d}. Consequently, a better understanding of the role of thermodynamics
 in extreme environments is needed. 
 
 Other processes such as 
 magnetic fields, feedback from protostars, and
intermittency of turbulence
may also be important and are not always considered in these models (see \S\ref{cluster_sec}  and the PPVI review by {\it Krumholz et al.}).
 For example, supersonic turbulence is intermittent and spatially correlated, so both velocity and density structures can exhibit coherency and non-Gaussian features, including significant deviations from log-normal density distributions \citep{FedRomKleetal2010,Hop2013b}. 
 In addition, the models above generally assume a steady-state system, where in fact time-dependent effects may be critical (see \S\ref{cluster_sec}). Certainly these effects are necessary to consider when translating the CMF models into predictions of star formation {\em rates} (\citealt{ClaKleBon2007,FedKle2012}; PPVI review by {\it Padoan et al.}). Preliminary attempts to incorporate time-dependence into the models \citep{HenCha2011,HenCha2013,Hop2013a} are worth exploring in more detail in future work.

\begin{figure}[t]
\begin{center}
 \includegraphics[width=8cm,angle=0]{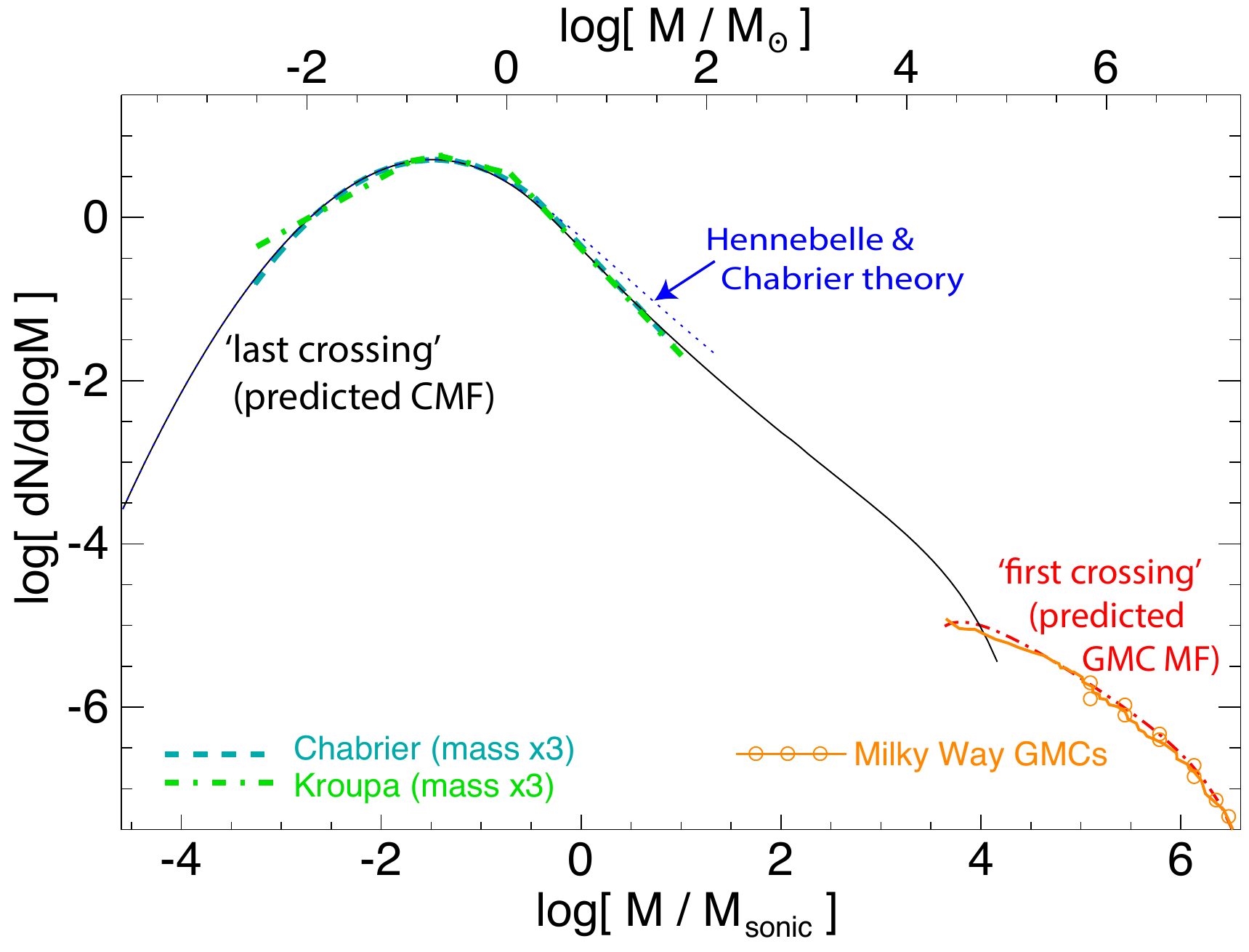}
 \caption{Core mass spectrum and GMC mass
spectrum calculated by \citep{Hop2012b} as a function of stellar mass and sonic mass. The grey dotted
curve shows the CMF inferred by \citep{HenCha2008}, 
while the 
lines indicate structures 
defined by the last (grey dashed) and first crossing (dot-dashed) of the self-gravitating
barrier. The \citet{Chabrier2005} and \citet{Kroupa2001} fits to the observed IMF are shown by the grey dashed and light-grey dot-dashed lines, respectively. Observed Milky Way GMCs 
are indicated by the solid grey line with circles.}
 \label{hopkins}
 \end{center}
\end{figure}

\subsection{The CMF Inferred from Numerical Simulations}

A variety of numerical simulations have been performed to study 
the origin of the CMF and the IMF (see \S\ref{cluster_sec} for a discussion of cluster simulations). 
Here, we consider a sub-set of numerical simulations that focus on modeling turbulence and the formation of cores.
The most important assumptions in these simulations are the explicit treatment
of self-gravity and turbulent driving, since these two processes lie at the very heart 
of the analytic models described above. 
Other physical processes such as magnetic fields and 
the equation of state are also very 
important but 
we temporarily ignore these for the purpose of comparing the analytic models to simulations.
A more technical but severe issue is the definition of the object mass used to 
compute the mass spectrum.  First, the exact definition of a core can be important, and 
second, ``sink'' particles are often  
 used to mimic protostars or cores that accrete from the surrounding gas. 
Given that these sinks cannot self-consistently model all the relevant physical processes, 
they are usually intermediate between a core and a star, although when feedback 
is properly accounted for, they may more closely resemble a star (see \S\ref{sec:interactions}). 
Another crucial issue is numerical resolution. 
Broadly speaking, the IMF covers about four orders of magnitude in mass, which 
corresponds to five to seven orders of magnitude in spatial scales 
assuming a typical mass-size relation $M_{_{\rm CORE}} \propto L^{1-2}$ (e.g., \citealt{Larson1981}). 
Since most numerical simulations span less than five orders of magnitude, the mass spectrum they compute is necessarily limited.

The CMF has been determined in 
 driven turbulence simulations without self-gravity, where
prestellar cores are identified as regions 
that would have been gravitationally bound if gravity 
had been included. 
The exact cores identified depend on the types of support
considered. 
\citet{PadNorKrietal2007} performed
both hydrodynamical and magnetohydrodynamical simulations without gravity. They found that 
when only thermal support was considered in the core determination, 
the mass spectrum was too steep with $\alpha = 3$. When 
magnetic pressure support was included in the 
core definition, the mass spectrum became shallower and $\alpha \simeq 2-2.5$,
which is similar to the Salpeter exponent.
\citet{Schmidtetal2010} performed similar simulations using both 
solenoidal and compressive random forcing and including 
thermal and turbulent support.
They also found that 
when only thermal support was used to assess core boundedness, the 
mass spectrum was too steep with $\alpha \simeq 3$.
When turbulent support was included, the mass 
spectrum again became shallower and the slope approached the Salpeter slope. 
\citet{Schmidtetal2010} also demonstrated that the position of the distribution peak noticeably depends
on the turbulent forcing. The peak shifts towards smaller masses for
compressive forcing.  This occurs because as the gas experiences more compression, smaller collapsing regions are generated; the sonic mass (and Jeans mass in those dense regions) became smaller because the fraction of compressive motions increased. 
This effect is also predicted by analytic models.
\citet{Schmidtetal2010}  performed  quantitative 
comparisons with both the predictions by 
\cite{HenCha2008} and the hydrodynamical 
predictions by \cite{PadNorJon1997}  and found good agreement.

Simulations that include both turbulence and self-gravity 
have been performed by a variety of groups (see \S\ref{cluster_sec}).
Of particular interest here, \cite{SmiClaBon2009} used simulations to  investigate the CMF, the IMF, and the 
correlation between them. 
They obtained the CMF by defining prestellar cores on the basis of either the
gravitational potential  or thermal and turbulent support and defined the IMF as the distribution of sink masses. 
\cite{SmiClaBon2009} found that both the simulated CMF and the particle mass function
resembled the observed IMF. To investigate the 
correlation between the two,  they 
identified the core in which a sink particle forms (the parent core), 
 and tracked the relationship between the sink 
masses and parent cores in time.
They found that the correlation remained very good for about three free-fall times, indicating that sink masses, at least  for early times, are determined by their parent core.
We discuss this work further and its relevance for understanding the link between the CMF and IMF in \S \ref{link_sec}.


\section{\textbf{THE LINK BETWEEN THE CMF AND THE IMF}}\label{link_sec}


Star forming molecular clouds are observed to contain ensembles of cores whose mass distribution closely resembles the stellar IMF. As discussed in \S \ref{cmf_sec}, theoretical models of turbulent fragmentation predict distributions of core masses that also resemble the IMF. At first glance, we would appear to be done: the theoretical core mass function resembles the observed core mass function, which in turn resembles the stellar IMF.

However, the picture is not so simple. If the shape of the IMF derives directly from the shape of the CMF, several conditions must hold. First, all observed cores must be truly ``prestellar", i.e., bound structures destined to condense into stars. Second, cores must not alter their mass by either accretion or mergers, or, if they do, they must do so in a self-similar fashion such that the shape of the CMF is preserved. Third, all cores must have the same star formation efficiency. Fourth, if cores undergo internal fragmentation, they must do so in a self-similar fashion. Fifth, all cores must condense into stars at the same rate; otherwise, cores that condense more slowly will be over-represented in the CMF. Figure \ref{fig:coretostar} illustrates what happens when each of these five conditions is violated. Finally, it remains to be shown that the cores produced by turbulent fragmentation theory can sensibly be identified with observed cores. \\

\begin{figure}
 \centerline{
 \includegraphics[width=3.3in]{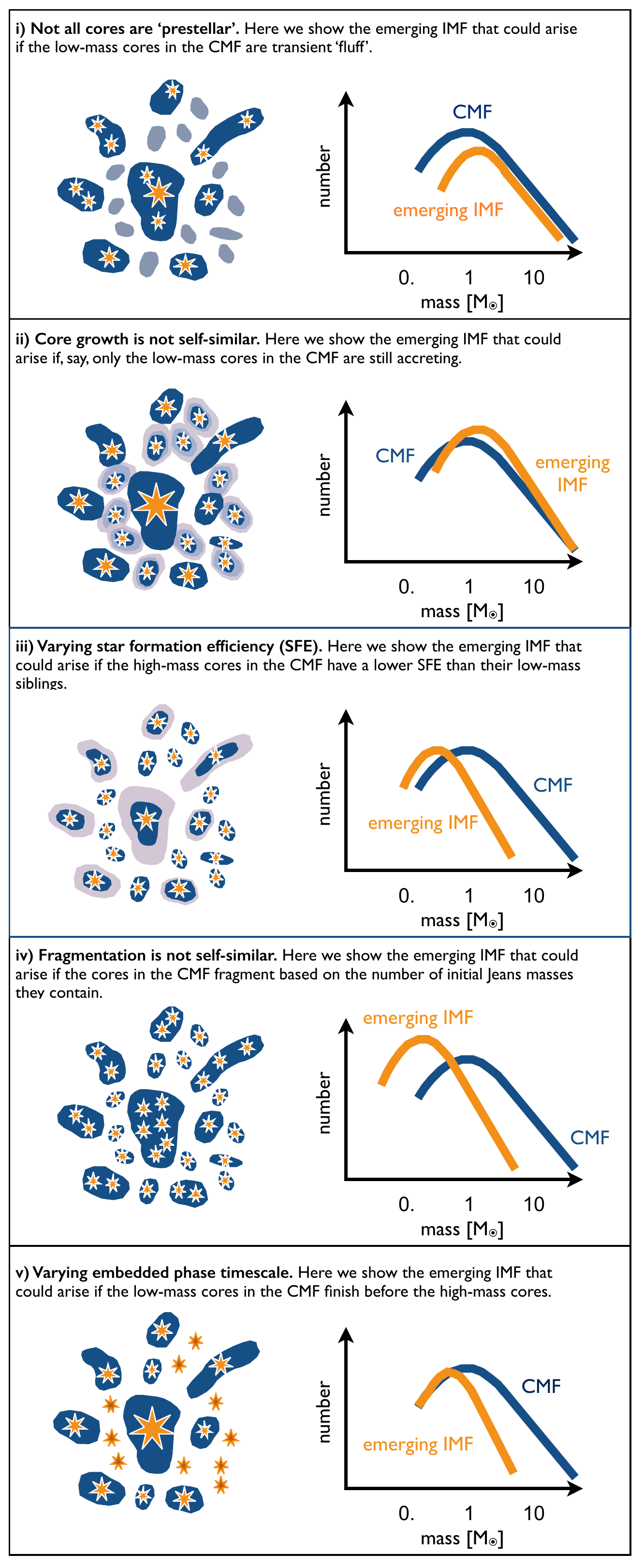}
 }
\caption{At the beginning of section \ref{link_sec}, we discuss the conditions that are necessary for the CMF to map to the IMF. These schematics illustrate what would happen if each of these conditions were to be violated.}
\label{fig:coretostar}

\end{figure}

\subsection{Inferring the Intrinsic Properties of Prestellar Cores}

\subsubsection{Masses and Thermal Support}

While observational studies of the CMF have tried to ensure that the cores identified are truly prestellar, there are many uncertainties. Since a prestellar core is by definition destined to form a single star or multiple system, an important quantity is the estimated ratio of the core mass to Jeans mass, which is commonly used by observers to determine whether a core is gravitationally bound. Unfortunately, this fundamental quantity is extremely sensitive to the derived temperature of the core, going as,
\begin{eqnarray}
\frac{M_{_{\rm CORE}}}{M_{_{\rm JEANS}}}\!&\!\sim\!&\!0.05\left(\frac{\pi}{\Omega^3}\right)^\frac{1}{4}\left(\frac{G\bar{m}F_{_\lambda}D}{c\kappa_{_\lambda}}\right)^\frac{3}{2}\frac{\lambda^6}{k_{_{\rm B}}^3T_{_{\rm DUST}}^3}.\hspace{0.5cm}
\end{eqnarray}
where $F_{_\lambda}$ is the monochromatic flux from the core, $D$ is its distance, $\Omega$ is the solid-angle enclosed by the core's boundary, $T_{_{\rm DUST}}$ is the dust temperature, $\kappa_{_\lambda}$ is the monochromatic mass-opacity (due to dust, but normalized per unit mass of dust {\it and} gas), $k_{_{\rm B}}$ is the Boltzmann constant, $G$ is the universal gravitational constant, $\bar{m}$ is the mean gas particle mass, and $c$ is the speed of light. 
Observational estimates of this ratio are very uncertain, primarily because $\kappa_{_\lambda}$ and $T_{_{\rm DUST}}$ are heavily degenerate \citep{SheKauSchetal2009b,SheKauSchetal2009a}. The situation has been improved by {\sc Herschel} \citep{AndMenBonetal2010}. By combining {\sc Herschel} multi-wavelength data with a careful statistical analysis, it is possible to break these degeneracies \citep{KelSheStuetal2012, LauStuSchetal2013}. However, estimates still typically assume that the dust and gas have the same temperature, which is only likely above a number density of $\sim 10^5$ cm$^{-3}$ when the dust and gas thermally couple (e.g., \citealt{GloCla2012}). Consequently, the question of whether an observed core truly belongs in the prestellar CMF (i.e.,  $M_{_{\rm CORE}}/M_{_{\rm JEANS}}>1$) can be fraught. 

\subsubsection{Non-thermal Support}

Constraints on the kinematics of cores can be obtained from observations of molecular line profiles (e.g., \citealt{AndBelMotetal2007}), but the kinematic information is normally convolved with the effects of inhomogeneous abundances and excitation conditions, finite resolution, superposition of emission from disparate regions along the line of sight, or optical-thickness (concealment of regions along the line of sight). In addition, spectra only probe the radial component of the velocity. These uncertainties aside, the line-widths measured from observations of N$_2$H$^+$, NH$_3$, and HCO$^+$ towards dense cores in Perseus and Ophiuchus suggest that the internal velocities, while generally sub-sonic, can contribute a significant amount of energy \citep{AndBelMotetal2007, JohRosTafetal2010, SchDadDiFetal2012}. 
Consequently, some of the cores identified purely on the basis of dust emission or extinction may actually be unbound,  transient objects.
 \citet{EnoEvaSaretal2008} demonstrated that this is likely to affect only the lowest mass-bins of their CMF, and then only by a small amount, but they note that the statistics are still small.

Similar constraints pertain to estimates of the magnetic field in cores. Field estimates are obtained using either the Zeeman effect, which only probes the line-of-sight component, or the Chandrasekhar-Fermi conjecture, which can be used to estimate the transverse component. \citet{CruHakTro2009} presented statistical arguments to suggest that magnetic fields are not able to support prestellar cores against gravity. 
However, observing magnetic field strengths in cores is challenging (see PPVI chapter by {\it Li et al.} for more discussion).

\subsubsection{Extracting Cores from Column Density Maps}

Any evaluation of the CMF inherently depends on the procedure used to identify cores and map their boundaries. Different algorithms (e.g., {\sc GaussClumps}, \citealt{StuGue1990}; {\sc ClumpFind}, \citealt{WildeGBli1994}; dendrograms, \citealt{RosPinKauetal2008}), even when applied to the same observations, do not always identify the same cores, and when they do, they sometimes assign widely different masses. Interestingly, different methods for extracting cores usually find similar CMFs even though there may be a poor correspondence between individual cores (for example, the CMFs derived for Ophiuchus by \citealt{MotAndNer1998} and by \citealt{JonWilMoretal2000}). A similar problem arises in the analysis of simulations \citep{SmiClaBon2008}. \citet{PinRosGoo2009} have shown that the number and properties of cores extracted often depend critically on the values of algorithmic parameters. Therefore, the very existence of the cores that contribute to an observed CMF should be viewed with caution, particularly at the low-mass end where the sample may be incomplete \citep{AndMenBonetal2010}.

\subsection{Phenomenology of Core Growth, Collapse and Fragmentation}
\label{core_growth_frag}

In the theory of turbulent fragmentation, cores form in layers assembled at the convergence of large-scale flows or in shells swept up by expanding nebulae such as H{\sc ii} regions, stellar wind bubbles and supernova remnants. Numerical simulations indicate that these cores are delivered by a complex interplay between shocks, thermal instability, self-gravity and magnetic fields,  moderated by thermal and chemical microphysics and radiation transfer \citep{BhaFraWatetal1998, Klessen2001, ClaBon2005, ClaBon2006, VazGomJapetal2007, PadNorKrietal2007, HenBanVazetal2008, BanVazHenetal2009, HeiNaaWal2011, WalWhiGir2012}. 
Due to the anisotropies introduced by shocks, magnetic fields, and 
self-gravity, these layers and shells tend to break up first into filaments from which cores then condense, in particular at the places where filaments intersect. However, all cores do not form stars; some disperse without forming stars and others may be assimilated by larger cores. 

The main focus of star formation is those cores that become gravitationally unstable. A number of authors have studied collapsing cores in numerical simulations \citep{KleHeiMac2000, Klessen2001, TilPud2004, ClaBon2005, ClaBon2006, VazGomJapetal2007, SmiClaBon2009}. These studies show that very low-mass objects ($M_{_{\rm CORE}}\!<\!0.5\,{\rm M}_{_\odot}$) tend not to be self-gravitating but are instead transient structures with excess turbulent and thermal energy.  
In contrast, the first gravitationally unstable cores to form in these simulations are always around the mean thermal Jeans mass, $M_{_{\rm CORE}}\!\sim\!{\rm 1M}_{_\odot}$. Structure decompositions of these clouds yield mass functions that resembles the CMF  even though most of the identified ``cores" are unbound and transient.

Studies that include sink particles and employ Lagrangian fluid methods have investigated the mapping between the masses of prestellar cores at the point they become bound and the final masses of the stars that form within them \citep{ClaBon2005, SmiClaBon2009}. To date, such studies have only been done with smoothed particle hydrodynamics, but a similar analysis could be performed using tracer particles in Eulerian fluid codes. As previously mentioned in \S\ref{cmf_sec} there appears to be a good correlation between core and star masses at early times, but the correlation erodes after several core dynamical times. \cite{ChabrierHennebelle2010} reanalyzed the \citet{SmiClaBon2009} data to quantify the CMF/IMF correlation by comparing the distributions with a simple Gaussian distribution. They found that the level of mismatch between stellar masses and the parent cores can be reproduced with a modest statistical dispersion of order one third of the core mass. 
However, the final mismatch between stellar masses and their natal core masses in these studies could in part be due to the dense initial conditions employed, the lack of feedback to halt accretion, and limited dynamic range. More recent simulations follow the assembly of cores from typical GMC densities (e.g., \citealt{GloCla2012}) or from even lower densities typical of the ISM (e.g., \citealt{HenBanVazetal2008, ClaGloKleetal2012}), and these may provide a more useful framework within which to evaluate the core formation process. Such studies have the advantage of capturing the transition between the line-cooling and dust-cooling regimes, and the characteristic stellar mass may depend upon the conditions of the gas at this transition \citep{Larson2005,JapKleLaretal2005}.

Numerical calculations suggest that the flow of material onto cores is highly anisotropic, where the largest and most fertile cores acquire much of their mass from the filaments in which they are embedded (e.g., \citealt{WhiChaBhaetal1995, SmiGloBonetal2011}). Moreover, some of the material flowing into a core may already have formed stars before it arrives  at the center \citep{GirFedBanetal2012}. These stars can then continue to grow by competitive accretion while cold, dense gas remains \citep{BonBatClaetal1997, DelClaBat2003} 
Stars may become detached from the gas reservoir if they are ejected through dynamical interactions with other stars \citep{ReiCla2001}. Even if the material flowing into a core has not yet condensed into a star when it arrives, the anisotropic inflow manifests itself as turbulence in the core, and therefore creates sub-structure that may help the core to fragment \citep{GooWhiWar2004,WalWhiGir2012}. 

However, early simulations and most of those above do not include MHD effects or radiative feedback (see also \S \ref{cluster_sec}).
When MHD \citep{TilPud2007, HenTey2008},
radiative feedback \citep{KruKleMcK2007, KruKleMcKetal2009, CunKleKruetal2011}, or both \citep{PriBat2007, ComHenHen2011, PetBanKleetal2011, MyeMcKCunetal2013} are included, the efficiency of fragmentation is reduced, and the number of protostars formed by a core may be reduced.
Magnetic torques may also act to redistribute angular momentum, which affects fragmentation \citep{BasMou1995}. 

When a protostar forms in a core it is normally attended by an accretion disk, which may fragment in some circumstances to form (usually low-mass) companions (e.g., \citealt{BonBat1994}). Early simulations indicated that this could provide an important channel for forming brown dwarfs and very low-mass H-burning stars \citep{BatBonBro2002,StaWhi2009a,StaWhi2009b}. MHD simulations have cast some doubt on whether sufficiently massive disks can form given observed mass-to-flux ratios, because magnetic torques  efficiently remove angular momentum. However, misalignment of the field relative to the angular momentum vector may provide an avenue for massive disks to form (e.g. \citealt{JooHenCia2012, SeiBanPudetal2012}). In addition, turbulence, Parker-like instabilities, and non-ideal effects may be important and must be included in simulations to resolve issue of massive disk formation.

Even if massive disks do form, it is critical to consider radiative feedback from the central star, and any other stars that form. If a star accretes continuously from its disk, it has a high accretion-luminosities, and the consequent radiative heating stabilizes its disk against fragmentation  (see \S\ref{cluster_sec}, PPVI review by {\it Krumholz et al.}). 
On the other hand, if the accretion is episodic, with most of the accretion energy being released in short bursts, and if the time between bursts is sufficiently long, then disk fragmentation can still occur and appears to produce low-mass stars in the numbers and with the statistical properties observed \citep{StaWhiHub2012}. \\

\subsection{High-Mass Cores and the High-Mass IMF}

\citet{McKTan2002} proposed that high-mass star formation occurs via the roughly monolithic collapse of high-mass cores in a scaled-up version of low-mass star formation. This implies that any mapping between the CMF and IMF must extend to high masses, as suggested by \citet{Oey2011}. A few $\simeq 10\, \msun$ prestellar cores have been reported in the literature and included in CMF determinations \citep{NutWar2007,AlvLomLad2007}, however, it is usually unclear whether these more massive objects will go on to form a single star system or are ``clumps", which will form a cluster of stars.
While several Class 0 candidates with masses greater than $50\, \msun$ have been found \citep{BonMotCseetal2010, PalFueGiretal2013, ZhaTanDeBetal2013}, there is no unequivocal  high-mass {\it prestellar} core. It is unclear whether this implies these objects never exist, and thus the CMF/IMF mapping does not extend to such high masses, or whether such cores are extremely short lived as proposed by \citet{HatFul2008}. 

Infra-red dark clouds (IRDCs) have often been suggested as the best candidates for high-mass prestellar cores given their high masses and column densities. However their clumpy interiors \citep{RagBerGut2009} and sub-virial velocity dispersions \citep{RagHeiBeretal2012} suggest that these objects  may be similar to local and lower mass regions of star formation such as Ophiuchus \citep{AndBelMotetal2007}. If so, these objects are more likely the progenitors of entire clusters \citep{RatJacSim2006}. More recent studies have shown that some parts of IRDC G035.39-00.33 may be in virial equilibrium \citep{HerTanKaietal2012} and thus consistent with the scenario proposed by \citet{McKTan2002}. Further interferometric studies of the interiors of these objects are needed and could be performed with ALMA.

It remains unclear whether the observed high-mass Class 0 candidates contain a single massive protostellar system or a deeply embedded protocluster, such as those discussed in \S\ref{cluster_sec}. Extensive radiative transfer modeling (along the lines of \citealt{ZhaTanMcK2013, OffRobHanetal2012}), may be able to help distinguish between these scenarios. Furthermore, Class 0 cores are not necessarily isolated objects; they are embedded within larger star forming regions. Observations of local regions suggest that the relative motion of low-mass cores is small (e.g., \citealt{AndBelMotetal2007}), however, in more clustered regions cores could be undergoing accretion and merging as described above. 

\subsection{Statistical Constraints}


As mentioned above, the statistical mapping of the CMF to the IMF requires that on average cores fragment self-similarly and have a roughly constant efficiency that is independent of mass. A simple self-similar mapping appears to be possible because a very simple model can reproduce binary statistics, such as binary frequencies and mass ratios, provided that the typical core forms 4 or 5 stars and has a high star formation efficiency \citep{HolWalGooetal2013}. A less restrictive mapping between the CMF and IMF would entail that the most abundant stars, $\overline{M} \!\sim\!0.2\,{\rm M}_{_\odot}$, mainly form 
from the most abundant prestellar cores, $\overline{M}_{_{\rm CORE}}\!\sim\!1\,{\rm M}_{_\odot}$, while the most massive stars form only from the most massive cores \citep{GooNutKroetal2008,Oey2011}. 

Initial estimates of the efficiency with which core gas forms stars are based on the premise that most cores produce a single star. Observationally, the CMF appears to have a peak at around 1 M$_{\odot}$ \citep[e.g.][]{AlvLomLad2007, NutWar2007, EnoEvaSaretal2008, AndMenBonetal2010}, which suggests an efficiency of $\sim\!30\%$. These early estimates have subsequently been increased to $\sim\!50\%$ to allow for the formation of binary systems. 
This estimate is consistent with the theoretical predictions of \citet{MatMcK2000}, who look at the impact of the generalized X-wind
model \citep{ShuNajOstetal1995} on a protostellar core. They predict
efficiencies in the range 20\% to 70\%, depending on the
geometry of the core, and stress that this effect should be insensitive
to the mass of both the protostar and its parent core. In addition,
efficiencies of around 50\% have been claimed in numerical
studies of the collapse of magnetized, rotating prestellar cores
\citep{MacInuMat2009, MacMat2012, PriTriBat2012}. 
These studies capture the Class 0/1 protostellar outflows that
arise both from the magnetic-tower structure, which dominates the
polar regions of the core \citep{Lyn1996}, and from the centrifugal
acceleration that is powered in the protostellar disk 
\citep{BlaPay1982, PudNor1983}.  While these results are
encouraging for the CMF-IMF mapping, it should be stressed that
both the theoretical and numerical predictions quoted here rely on
idealized geometries; the final efficiencies in more realistic,
turbulent cores, such as those studied by \citealt{SeiBanPudetal2012}, are
still unknown. Finally,  
\citet{HolWalGooetal2013} argued that the efficiency might be much higher and perhaps even larger than 100\% if cores continue accreting while forming stars.

Although efficiencies must be taken into account when mapping the CMF to the IMF, the observed CMF peak is also quite uncertain. CMF peaks have been reported to be $\sim \,0.02\, \msun$ in Polaris \citep{AndMenBonetal2010}, $\sim \,0.1\, \msun$  in $\rho$-Oph \citep{MotAndNer1998}, $\simeq \, 2\, \msun$ in the Pipe Nebula \citep{AlvLomLad2007},  and $\sim \, 1\, \msun$ in Orion, Aquilla and the combined populations of $\rho$-Oph, Serpens and Perseus \citep{NutWar2007,EnoEvaSaretal2008,AndMenBonetal2010}.  \citet{GooNutKroetal2008} noted that the peak of the CMF appears to depend on the source distance, which could occur if blending creates bias in more distant regions. As discussed in \S3.1, the different peak values may be due to different core definitions (e.g., the inclusion of low-density unbound objects), but they may also due to observational incompleteness, which is often near the peak position.

Additional constraints on the CMF-IMF mapping are imposed by considering the time evolution of the CMF. For example, \citet{ClaKleBon2007} stressed that cores must have at least a Jeans mass to form stars. Assuming a roughly constant temperature, this presents two options: {\it either} the cores all have the same Jeans mass ($\sim 0.1\, \rm M_{\odot}$), and so cores contain different numbers of Jeans masses; {\it or} the cores each have a few Jeans masses such that the low-mass cores are denser than high-mass cores. The latter scenario is more likely to result in the self-similar mapping between the CMF and IMF discussed above, since it avoids the problem of an extremely Jeans-unstable core fragmenting to form a large cluster. However, this scenario also implies that cores likely evolve on different timescales, and low-mass cores form protostars more quickly than their high-mass counterparts. 
The emerging stellar mass function should therefore be biased towards lower masses than the full IMF, which is not evident in nearby regions. If different mass stars form on different timescales, \citet{McKOff2010} showed that the present-day protostellar mass function will be distinct from one in which stars form on the same timescale. Since the masses of embedded protostars are difficult if not impossible to measure, the distribution of protostellar luminosities provides a promising indirect method to constrain the underlying star formation times and accretion rates \citep{OffMcK2011}.

The observational evidence for differences between prestellar and protostellar core mass distributions is ambiguous. \citet{HatFul2008} showed that the mass distribution of prestellar and protostellar cores in Perseus were markedly different: the former resembles the IMF, while the latter is strongly peaked at around 8 M$_{\odot}$. 
One interpretation of these observations is that high-mass cores evolve {\em faster} than low-mass cores. Given the criterion for Jeans instability, it is not obvious how this could be the case. Alternatively, \citet{HatFul2008} noted that these results could imply that most low-mass cores are transient. As discussed above, such a picture is broadly consistent with what is found in simulations and would imply that the observed CMF is seriously contaminated by unbound clumps. Such a scenario has also been used to explain the chemistry of low-mass cores in L673 \citep{MorGirEst2005}. In contrast, \citet{EnoEvaSaretal2008} measured the combined CMF of $\rho$-Oph, Serpens and Perseus and demonstrated that the prestellar and protostellar CMFs have statistically similar distributions.  Since the \citet{HatFul2008} sample contains less than half the number of cores of the \citet{EnoEvaSaretal2008} sample, the apparent differences may be purely statistical (see also the PPVI chapter by {\it Dunham et al.} for discussion of the challenges of observing and classifying protostars). 

The Press-Schechter and excursion-set IMF models discussed in \S\ref{cmf_sec} 
potentially provide a solution to the timescale issue. The models
envisage that low-mass cores are more likely to be disrupted by
turbulence than high-mass cores -- an effect that \citet{HenCha2008} refer
to as ``turbulent dispersion''. This process acts to lower the overall
star formation rate in the low-mass cores, thus helping to
counteract the fact that low-mass cores may, individually, collapse
faster than their higher-mass counterparts. Again, this would imply
that not all the observed cores are truly ``prestellar''. 
They also find that the cores that make up the bound CMF have similar densities yet still follow an IMF-like distribution. While this potentially provides a  resolution to the apparent timescale problem, it does not address the question of why a 10 M$_{\odot}$ core containing many Jeans masses would only form a small number of objects. One solution to this puzzle may be supplied by the addition of more complex physics, which is discussed further in the following section.

\section{\textbf{THE FORMATION OF CLUSTERS}}\label{cluster_sec}

 
In our Galaxy, many stars are observed to form in groups, associations, or clusters \citep{LadLad2003,BreBasGutetal2010}, and the IMF is usually measured by observing these systems because they provide ensembles of stars with similar ages.  Thus far, our theoretical discussion has focussed on the dense cores that are observed to be the sites of star formation, tackling their formation, properties, and whether and how they produce stars.  We have largely considered them to evolve in isolation from each other with each producing one or perhaps a few stars.  However, particularly in dense star-forming regions, the formation of a cluster and the formation of stars occurs simultaneously.  Thus, the formation of stellar clusters is a dynamical and time-dependent process, where protostars may interact with each other and the evolving cloud via many different physical mechanisms.  Due to this time dependence and inherent complexity, numerical simulations are essential to explore the role of each mechanism in producing the IMF.

In the mid-1990s, interest in ``competitive accretion" as a mechanism for producing a mass spectrum was revived by studies using smoothed particle hydrodynamics (SPH) with sink particles representing protostars 
\citep{BonBatClaetal1997}.  Simulations of protostellar ``seeds" embedded in and accreting from a gas reservoir demonstrated that the initial mass spectrum was relatively unimportant \cite[see also][]{Zinnecker1982}, but that differential accretion naturally produced both a mass spectrum and mass segregation, with the more massive stars located near the cluster center as is often observed in young clusters. These early studies identified the importance of the stellar spatial distribution and dynamical interactions between protostars. In some cases, dynamical interactions were able to eject stars from the cloud and thus halt accretion.  Subsequent papers found that the high-mass end of the mass spectrum was steeper than ${\rm d}N/{\rm d} \log M \approx M^{-1}$ due to changes in the dynamics of accretion and mass segregation \citep{BonBatClaetal2001,BonClaBatetal2001}.  Rather than begin with protostellar ``seeds", \cite{KleBurBat1998} studied the fragmentation of structured gas clouds to form protostars and followed their subsequent competitive accretion.  They found a log-normal protostellar mass spectrum resembling the observed IMF  and again noted the role of dynamical interactions in terminating accretion.  

These early studies laid the foundation for the research performed over the last decade, which has included increasingly more physical processes.  In this section we review numerical work investigating the roles of initial conditions, protostellar interactions, and feedback on the IMF.

 \subsection{Role of Initial Conditions in Shaping the IMF}
 \label{sec:initialconditions}
 
 A variety of different initial conditions and physical effects have been explored in simulations of star cluster formation. A number of these find good agreement with the observed IMF, while others highlight interesting variations. However, we stress that currently no simulations of cluster formation include all the important physics, produce sufficient stellar statistics, {\it and} reach sufficiently high-resolution to resolve close binaries.  Moreover, simulations that follow the accretion and dynamical evolution of stars employ sink particles as a numerical convenience  in order to limit the resolution of gravitational collapse and reduce computational expense. Consequently, the results summarized here are best considered suggestive ``numerical experiments."

 \subsubsection{Cloud Initialization}
 
 Simulations typically adopt one of three main approaches to the problem of cloud initialization. The first approach is to consider an isolated cloud, typically with an initially analytic density distribution (e.g., 
\citealt{GirFedBanetal2011}). Cloud structure may be seeded by applying a random velocity field at the initial time. A second approach is to consider a piece of a molecular cloud by adopting periodic boundary conditions (e.g., \citealt{KleBur2000,OffKleMcK2008}). In the latter case, the large-scale turbulent energy cascade can be modeled by continuous turbulent driving \citep{Maclow1999}.   Both of these approaches dispense with the issue of molecular cloud formation and instead initialize with mean observed molecular cloud properties. Both of these approaches produce a CMF and/or IMF similar to observations (e.g., \citealt{Bate2012,KruKleMcK2012}, see Figure \ref{cdf}), which suggests the IMF may be fairly insensitive to the details of cloud formation in numerical simulations. Another approach directly models the cloud formation. For example, some simulations begin with the collision of two opposing atomic streams of gas   \citep{KoyInu2002,VazGomJapetal2007, HeiHarBur2008, HenBanVazetal2008,BanVazHenetal2009}.   
 This setup is predicated on the idea that some molecular clouds form from coherent gas streams  (see the PPVI chapter by {\it Dobbs et al.} for further discussion of cloud formation mechanisms). 
 An additional promising approach begins with a full or partial galactic disk that forms clouds self-consistently by including ISM physics. While this achieves more realistic and self-consistent initial cloud conditions, current resolution is not sufficient to follow the formation of individual stars.

\subsubsection{Turbulent Properties}

Most cloud simulations include some degree of supersonic turbulence. 
The details of this turbulence have been widely varied, reflecting both current observational uncertainties and underlying theoretical disagreement. Over the last decade simulations have typically adopted one of two treatments. Either turbulent motions are continually driven to maintain a fixed turbulent energy or the turbulence is allowed to freely decay after initialization. Arguments in favor of the former appeal to the very low star formation rate in clouds  \citep{KruTan2007}, the continuing cascade of energy from larger scales, and the approximate gravitational and kinetic energy equipartition in clouds \citep{KruMatMcK2006}. However, approaches applying random velocity perturbations in lieu of kinematic stellar feedback are not physical (see \S5.3). Alternatively, some turbulence may be regenerated through gravitational collapse \citep{HufSta2007, FieBlaKet2008, KleHen2010, RobGol2012}. Both approaches have been successful in reproducing IMF characteristics (e.g., see  \S\ref{cmf_sec}, \citealt{Bate2012, KruKleMcK2012}). 
Clouds with excess turbulent motions, i.e., gravitationally unbound clouds, have lower star formation efficiencies and produce IMFs that are flatter than observed \citep{ClaBonKle2008}.

The turbulent outer driving scale is generally assumed to be comparable to the simulated cloud size on the basis that turbulence is generated during the cloud formation process and cascades smoothly to small scales (but see also \citealt{SwiWel2008}). \citet{Klessen2001} demonstrated that driven turbulence injected on small scales produces a CMF/IMF that is too flat compared to observations. In the case of decaying turbulence, variations in the turbulent power spectrum slope 
apparently have little effect on the IMF \citep{Bate2009c}. \citet{GirFedBanetal2011} found that the resultant IMFs were also relatively insensitive to whether the initial turbulence was solenoidal (divergence-free), compressive, or some combination of the two (see Figure \ref{cdf}). 

One universal characteristic of turbulent simulations, independent of turbulence details, is the production of filamentary structure (e.g., \citealt{HeiHarSlyetal2008, ColKriPadetal2012}). 
Star-forming cores often appear at regular intervals along filaments \citep{MenAndDidetal2010, OffKruKleetal2008, SmiSheStuetal2012}, and gas may flow freely along them, supplying additional material to forming stars (see PPVI review by Andr{\'e} et al. for additional discussion of filaments). However, it remains to be seen whether observed filamentary structure is simply a consequence of turbulence or serves some more fundamental role in the determination of the IMF.

\subsubsection{Initial Density Profiles}

Simulations have explored a range of underlying analytic density profiles, e.g.,  $\rho( r) \propto r^{-\beta}$ with $\beta$ ranging from 0 to 2.
Most recently, \citet{GirFedBanetal2011}  and \citet{GirFedBanetal2012}  found that turbulent, centrally condensed density profiles form a more massive star and fewer lower mass stars, whereas turbulent, flat density profiles produce more distributed clusters. The different density profiles also yielded demonstratively different IMFs, where the IMFs produced by uniform and Bonnor-Ebert density distributions more closely resembled the observed IMF (see Figure \ref{cdf}).  

\subsubsection{Magnetic Field Strength}

The degree of magnetization and the relative importance of turbulent 
and magnetic energy on different scales remain controversial, but there is consensus that magnetic support can have a dramatic impact on molecular cloud evolution. Strong magnetization provides pressure support that reduces the core and star formation rate \citep{HeiMacKles2001, PriBat2009, DibHenPinetal2010, PadNor2011a, PadHauNor2012}. \citet{LiWanAbeetal2010} showed that magnetic fields can also reduce the characteristic stellar mass by increasing the number of low-mass stars formed in filaments and reducing the number of intermediate mass stars formed from  turbulent compressions. The authors attributed this difference to the suppression of intermediate mass stars from converging flows; magnetic support inhibited isolated collapse and instead stars formed predominantly within magnetized filaments, which tended to produce smaller mass stars. An anti-correlation between characteristic stellar mass and magnetic field strength was also demonstrated in simulations of star formation in the galactic center \citep{HocSchSpaetal2012}.  In principle, a non-ideal MHD treatment would allow the field to diffuse from magnetically subcritical regions, however, few simulations have included such effects and the impact on the shape of the IMF is unclear. \citet{VazBanGometal2011} found that including ambipolar diffusion had little effect on the star formation rate in their simulations.  Larger cloud simulations that produce more stars and include non-ideal effects are necessary to fully understand the effect of magnetic fields on the IMF.

\subsubsection{Thermal Assumptions}

The role of thermal physics in shaping the IMF remains an open question. If the characteristic stellar mass is due to turbulent fragmentation then thermal physics may play only a minor role (see \S\ref{cmf_sec}). Alternatively, thermal physics, by setting the characteristic Jeans mass, may set the peak of the IMF \citep{JapKleLaretal2005, Larson2005, BatBon2005, BonClaBat2006, KleSpaJap2007}. Simulations have shown that the peak of the IMF is sensitive to the adiabatic index, $\gamma$, of the gas and the critical density at which the gas transitions from being isothermal (efficient cooling) to adiabatic (inefficient cooling). For a polytropic equation of state, $P=K \rho^{\gamma}$, the characteristic Jeans mass can be expressed as $M_{\rm J} = \pi^{5/2}/6 (K/G)^{3/2}\gamma^{3/2}\rho^{(3/2)\gamma-2}$, where $\rho$ is the gas density, $K$ is a constant, $G$ is the gravitational constant \citep{JapKleLaretal2005}.  Proponents of this model argue that while turbulent and magnetic support may delay gravitational collapse such support will eventually dissipate or diffuse leaving thermal pressure alone to oppose gravity. In either case, the peak cannot be too sensitive to the cloud environment without producing large variations, which is hard to reconcile with observations of a generally invariant IMF.   However, if the gas can be described by a barotropic equation of state  with some fixed critical density, as suggested by \citet{Larson2005}, the IMF peak mass may be set independently of the cloud Jeans mass and, thus, partially decouple the IMF from the cloud environment  
\citep{BonClaBat2006}. If only thermal physics is important and if the effective equation of state depends only on fundamental atomic parameters and does not also depend strongly on metallicity then this could explain an invariant IMF.

More detailed thermal physics can be considered by including heating and cooling by dust, molecules and atomic lines in place of a fixed equation of state. \citet{MyeKruKleetal2011} found that simulations including radiative transfer produced only weak variations in the IMF even as the metallicity varied by a factor of 100. 
Using a different approach, \citet{GloCla2012} performed simulations including heating and cooling computed via a reduced chemical network. These showed that the star formation rate is also relatively insensitive to  the gas metallicity for variations over a factor of 100.
Similarly, \citet{DopGloClaetal2011} found that dust cooling enabled small-scale fragmentation and hence the formation of low-mass stars for $Z \ge 10^{-5}\, {\rm Z}_{\odot}$.  Altogether, these studies help explain the observed IMF invariance over environments with a range of metallicities. 


\begin{figure*}[t]
\epsscale{1.4}
\plotone{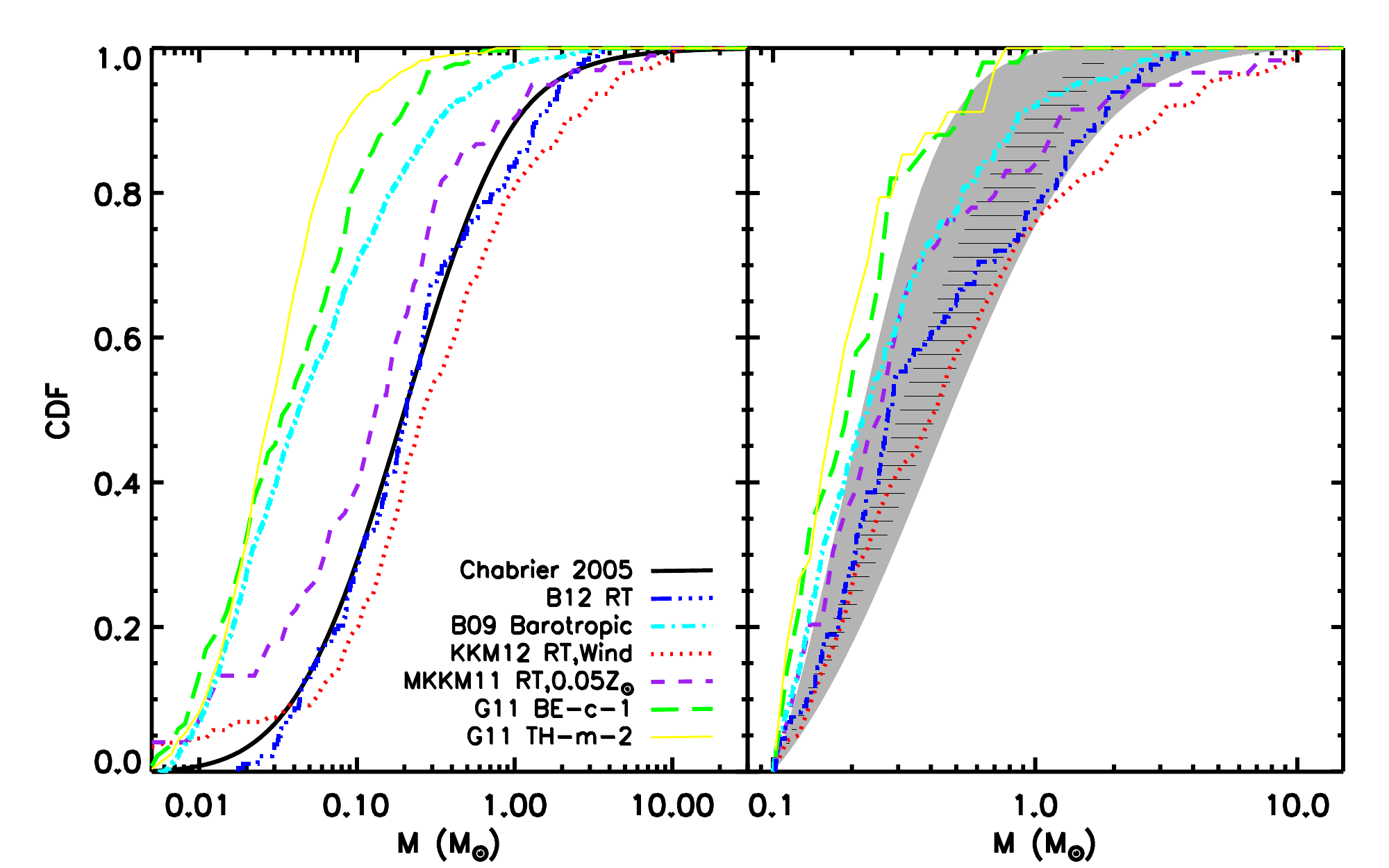}
\vspace{-0.1in}
\caption{Cumulative distribution functions (CDFs) for stellar masses from the simulations BE-c-1 (Bonnor-Ebert sphere with decaying compressive turbulence), TH-m-2 (uniform density with decaying mixed turbulence) both from \citet{GirFedBanetal2011}, simulation with 0.05\,Z$_{\odot}$ and radiative feedback \citep{MyeKruKleetal2011}, simulation with radiative feedback and protostellar outflows \citep{KruKleMcK2012}, simulation with decaying turbulence and a barotropic EOS \citep{Bate2009a}, and simulation similar to B09 but including radiative transfer \citep{Bate2012}. Left: The CDFs are shown with the \citet{Chabrier2005} single-star IMF for a mass range of 0.005-100 $\msun$. Right: The same CDFs but computed only for the mass range 0.1-15\,$\msun$. The shaded regions show the fits and associated error for the observed IMF at the half-mass cluster radius for Westerlund 1 (solid grey, {\it Andersen et al.~in prep}) and for the combined data of the Pleiades \citep{MorBouStaetal2003}, IC4665 \citep{deWBouPaletal2006} and Blanco 1 (\citealt{MorBouStaetal2007}, horizontal lines). The fits of the observational data are performed for the ranges 0.15 -15\,$\msun$ and 0.045-2\,$\msun$, respectively. 
\label{cdf}}
\end{figure*}

\subsection{Role of Protostellar Feedback and Interactions}
\label{sec:interactions}

It's possible that the protostars themselves play a part in setting the IMF since they may interact with each other and the molecular cloud as the cluster forms.  If they play the dominant role, this may explain why the IMF is not observed to vary greatly with environment (see \S \ref{obs_sec}); the generation of the IMF may be a self-regulating process rather than depending sensitively on initial conditions.

\subsubsection{Dynamical Interactions}

There are many ways by which protostars may affect the cloud in which they form and each other.  As mentioned previously, dynamical interactions between accreting protostars can produce a spectrum of masses because the accretion rate of any particular protostar depends on its location, the surrounding gas density, and its speed.  Typically, slow-moving protostars located near the cloud centre, where the overall gravitational potential of the cloud tends to funnel the gas, accrete at higher rates than those in the outskirts of the cluster or those that have been involved in dynamical interactions that have given them higher speeds.  However, the accretion rate of any particular protostar also tends to be highly variable \citep{BonBatClaetal1997, Klessen2001}.  Dynamical interactions can be an effective mechanism for stopping accretion (by ejecting protostars from the cloud or increasing their speeds so that their accretion rate becomes negligible), and theories of the IMF can be constructed based on accreting protostars with probabilistic stopping times \citep{BasJon2004, BatBon2005, Myers2009} in which accretion may be terminated due to dynamical interactions, outflow or radiative feedback, and other mechanisms.  As mentioned in Section \ref{sec:initialconditions}, however, hydrodynamical simulations of competitive accretion alone result in IMFs whose characteristic stellar mass depend on the initial Jeans mass in the cloud \citep{BatBon2005, JapKleLaretal2005, BonClaBat2006} and typically produce too many brown dwarfs \citep{Bate2009a}.  \cite{BatBon2005} argue that the dependence on the initial Jeans mass comes about within competitive accretion models because the characteristic stellar mass is given by the the product of the typical accretion rate (which scales as $c_{\rm s}^3/G$, where $c_{\rm s}$ is the sound speed) and the typical dynamical interaction timescale which terminates accretion (scales with density as  $\rho^{-1/2}$), thus resulting in a scaling.  This is problematic because the IMF is not observed to vary greatly with initial conditions.

\subsubsection{Radiative Heating}

The effects of radiation transfer and radiative feedback from protostars has been investigated in star cluster calculations following three different approaches
\citep{Bate2009b, OffKleMcKetal2009, UrbMarEva2010}.  These studies found that the inclusion of radiative heating in the vicinity of protostars dramatically reduced the amount of fragmentation, particularly of massive circumstellar disks.  In turn this reduced the rate of dynamical interactions between protostars.  Since in the competitive accretion process, brown dwarfs and low-mass stars are usually formed as protostars that have their accretion terminated soon after they have formed by dynamical interactions, radiative feedback also reduced the proportion of brown dwarfs to stars. Figure \ref{cdf} illustrates that an over-abundance of low-mass objects is one of the main areas of disagreement between some simulations and observations. Potentially of even more importance, however, was the discovery that radiative feedback also removed the dependence of the mass spectrum on the initial Jeans mass of the clouds \citep{Bate2009b}.  \cite{Bate2009b} presented a qualitative analytical argument for how radiative feedback weakens the dependence of the IMF on the initial conditions.   For example, consider two clouds with different initial densities.  In the absence of protostellar heating, the Jeans mass and length are both smaller in the high density cloud, so fragmentation will tend to produce more stars, which have smaller masses and are closer together, than in the low-density cloud.  However, protostellar heating is most effective on small scales, so when protostellar heating is included in the higher density cloud, it raises the effective Jeans mass and length by a larger factor (resulting in fewer stars with greater masses) than when it is included in the lower density cloud (where the stars would already form further apart).  Thus, the dependence on the initial density is weakened.  More recently \cite{Krumholz2011} took this argument further to link the characteristic mass of the IMF with fundamental constants.  Regardless of whether these analytic arguments portray the full picture or not, subsequent radiation hydrodynamical simulations have shown that excellent agreement with the observed IMF can be obtained, along with many other statistical properties of stellar systems \citep{Bate2012}.

On the other hand, in very dense molecular clouds that produce massive stars, radiative feedback can be too effective at inhibiting fragmentation.  \cite{KruKleMcK2011} showed that radiative feedback could lead to an overheating problem in which, after an initial cluster of protostars is formed, further fragmentation is prevented while the protostars continue to accrete.  In this case, the characteristic stellar mass continually increases in time, leading to a top-heavy IMF.  When combined with a strong magnetic field (mass to flux ratio $\sim 2$), \cite{ComHenHen2011} and \cite{MyeMcKCunetal2013} both demonstrate that radiative feedback and magnetic support could almost completely prevent fragmentation of such initial conditions, leading to a single massive star or binary rather than a cluster. However, the details of fragmentation appear to be sensitive to a combination of the initial cloud conditions and the code applied. For example, when modeling the collapse of high-mass cores \citet{HenComJooetal2011} found that the effects of radiation in reducing fragmentation are modest in comparison to magnetic fields. Likewise, work investigating radiative heating, including ionization from high-mass sources, and using a long-characteristics radiative transfer approach, found a more modest impact by radiation on fragmentation \citep{PetBanKleetal2010,PetKleMacetal2010, PetBanKleetal2011}. In particular, \citet{PetKleMacetal2010} modeled a 1000\,$\msun$ clump, finding that fragmentation was mainly suppressed in the inner $\sim$1000 AU and otherwise continued unabated within the dense filaments feeding the central region. In any case, current 2D and 3D radiation-hydrodynamic methods adopt approximations of the radiative transfer equations. To better understand the effects of radiative feedback on the IMF in clusters, more work is required to implement methods that accurately follow the propagation of multi-frequency, angle-dependent radiation and achieve accuracy in both the optically thick and thin limits.

\subsubsection{Kinematic Feedback: Outflows and Ionization}

Another form of feedback is kinetic feedback, via protostellar jets and outflows, stellar winds, and supernovae.  All of these likely contribute to the turbulent motions within clouds and affect the efficiency of star formation \citep[e.g.][]{LiNak2006}, but the level of the different contributions is unclear (see the PPVI chapter by Krumholz et al.).  Calculations to determine how kinetic feedback may contribute to the IMF have only started to be performed recently, and in all cases to date the outflows have been added in a prescribed manner rather than generated self-consistently. \cite{DalBon2008} investigated the effects of stellar winds in hydrodynamical simulations of cluster formation.  They found that outflows slowed the star formation process, but had little effect on the IMF except at the high-mass end where they disrupted the accretion of high-mass protostars.  \cite{WanLiAbeetal2010} performed MHD simulations with and without outflows and came to the same conclusion -- outflows slowed the star formation rate (by a factor of two) and reduced the accretion rates of the most massive stars by disrupting the filaments that feed them and slowing the global collapse of the cloud.  \cite{LiWanAbeetal2010} investigated the effects of magnetic fields and outflows on the IMF, finding that although the general forms of the mass spectrums were the same in all calculations, strong magnetic fields reduced the characteristic stellar mass by a factor of four over hydrodynamical simulations, and that adding outflows to the MHD simulation decreased the characteristic mass by another factor of two. \citet{HanKleMcKetal2012} performed the first cluster formation calculations to include both radiative feedback and outflows.  Although only one third of the material accreted by a protostar was ejected in each outflow, the characteristic mass of the stars was reduced by two thirds relative to calculations without outflows.  They also pointed out that because the protostellar accretion rates were lower, radiative feedback was less important.  \cite{KruKleMcK2012} expanded on this, showing that the combination of outflows and turbulence reduced the overheating problem.  The protostellar population produced by their calculations with turbulence and outflows was in statistical agreement with the observed IMF.




Finally, ionizing feedback by massive stars can both trigger fragmentation by compressing dense cores and suppress star formation by destroying dense molecular gas and disrupting accretion flows \cite{DalBonClaetal2005,DalClaBon2007,WalWhiBisetal2013}.  Thus far, however, no discernible effect on the IMF has been found \citep{DalBon2012}.  However, it is important to note that the ionizing radiation included in these calculations only determines whether the gas is ionized and hot ($T \approx 10^4$~K) or molecular and cold ( $T \approx 10$~K).  Heating of the cold molecular gas by non-ionizing photons is not included.  Its inclusion may lead to a change in the IMF similar to that discussed in the previous section.   \citet{PetKleMacetal2010} do include ionization and heating by non-ionizing photons but do not make a prediction for the IMF.

\subsection {Other Predictions and Observables}

\subsubsection{Gas Kinematics}

At early times, protostars are often too dim and too enshrouded by dust and gas to make observational determinations about their spatial distribution, velocities and masses. However,  observations of dense gas kinematics provide spatial and dynamical clues about the initial conditions and properties of protostars. For example, the turbulent  line widths of low-mass cores are approximately sonic, where protostellar cores do not possess significantly broader widths than starless cores \citep{KirJohTaf2007, AndBelMotetal2007, RosPinFosetal2008}. In addition, the relative velocity offsets between the dense gas and core envelopes are small, suggesting that protostars form from gas that does not move ballistically with respect to its surroundings \citep{KirJohTaf2007,KirPinJohetal2010, AndBelMotetal2007}, which disfavors a dynamical scenario at very early times.  

Simulations can be compared directly to these metrics through synthetic observations
using radiative transfer.  Both \citet{AylLanCohetal2007} and \citet{OffKruKleetal2008}  found reasonable agreement between simulated and observed line widths for different numerical parameters. \citet{OffHanKru2009} showed that protostellar dispersions in simulations of turbulent fragmentation are initially small and motions are well-correlated with the surrounding gas. \citet{RunHarAcretal2010} verified this for star-forming cores in simulations where protostars later underwent competitive accretion. Molecular line modeling by \citet{SmiSheStuetal2012} and \citet{SmiSheBauetal2013} explored the evolution of line profiles for higher mass cores and filaments, where observations are often messier and more difficult to interpret than for local low-mass star forming cores. Additional work compares simulated dust continuum observations of forming protostars with observations \citep{KurHarBatetal2004, OffRobHanetal2012}. Such detailed synthetic observations are needed to constrain protostellar masses, gas morphologies, and outflow characteristics, which are essential metrics for distinguishing between IMF theories.  Future work requires additional quantitative synthetic observations in order to provide more discriminating observational comparisons.


\subsubsection{Cluster Evolution and Mass Segregation}

Stars may significantly dynamically interact during and after
formation. Understanding their initial spatial distribution and subsequent
evolution is necessary to accurately account for dynamical effects when
measuring the IMF in clusters. The stellar spatial distribution may also present some clues for distinguishing between different theoretical models.  For example, in the competitive accretion picture the most massive stars form in a clustered environment and are naturally mass segregated at birth \citep{BonBatClaetal2001}. In this picture instances of isolated massive stars are generally  stars ejected from a cluster via dynamical interactions. In the turbulent core model, the massive cores that are predicted to form massive star systems are more likely to be found in the cloud center, where the highest column density gas is located \citep{KruMcK2008}. However, massive stars are not physically precluded from forming in isolation \citep{TanKonButetal2013}. The observational evidence for isolated massive star formation is obfuscated by small number statistics and the large distances to massive targets. However, young, isolated massive stars appear to exist. \citet{BreBasEvaetal2012} and \citet{OeyLamKusetal2013} located a number of isolated massive stars that do not appear to have been dynamically ejected from a cluster. \citet{LamOeyWeretal2010} found that a couple of apparently isolated massive stars are in fact runaways, while several more appear to be in very sparse clusters. Using a sample of 22 candidate runaway O-stars, \citet{deWTesPaletal2005} concluded that only $\sim 4$\% of Galactic O-type stars form outside of clusters. Explaining the frequency of isolated massive stars and the mass distributions of small clusters is an ongoing challenge for theoretical models. 




\begin{figure}[t]
\vspace{-0.1in}
\epsscale{1.0}
\plotone{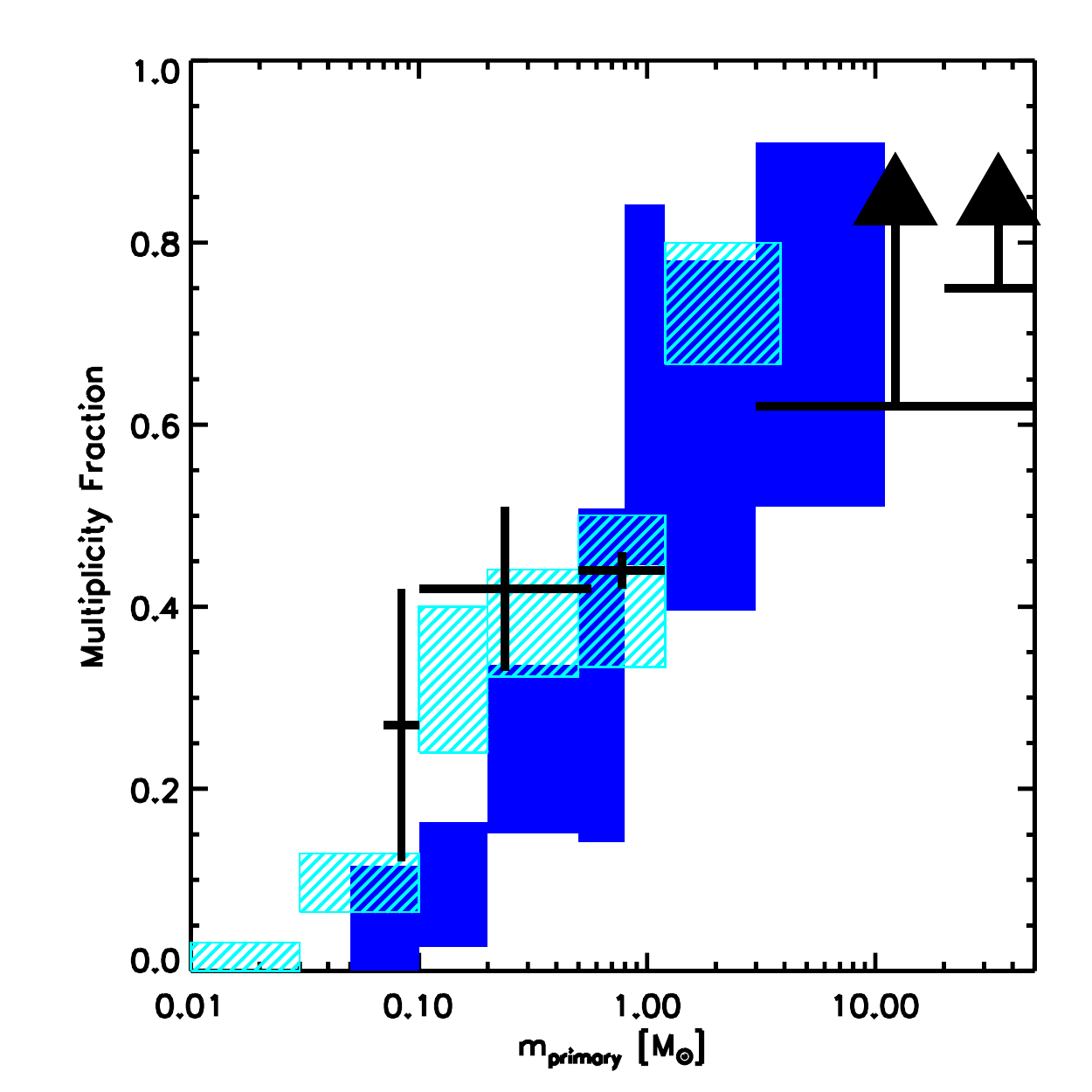}
\vspace{-0.2in}
\caption{Multiplicity fraction as a function of primary stellar mass. The solid shaded blocks indicate the multiplicity and associated statistical uncertainty for a simulation with radiative and kinematic feedback from forming protostars \citep{KruKleMcK2012}. The striped blocks show the multiplicity and statistical uncertainty for a simulation with radiative transfer and decaying turbulence \citep{Bate2012}. The black crosses represent observational results, with the horizontal width indicating the mass range for the observations and the vertical range showing the stated uncertainty. The two highest mass observational data points are lower limits. 
The data shown are taken from, from left to right, \citet{BasRei2006} and \citet{Allen2007} (shown as a single combined point), \citet{FisMar1992}, \citet{RagMcAHenetal2010}, \citet{PreBalHofetal1999}, and \citet{MasHarGieetal2009}.
\label{mult}}
\end{figure}

\subsubsection{The Multiplicity of Protostellar and Stellar Systems}

One important, intermediate step between the core mass function and single star IMF is the frequency and mass dependence of multiple star systems. Resolving close binary systems and accurately counting wide binary systems remains an observational challenge. However, there is convincing evidence that younger systems have higher multiplicity than field stars \citep{DucKra2013}. This is not surprising since higher order multiple star systems naturally dynamically decay with time.  As shown in Figure \ref{mult}, simulations with sufficient resolution have successfully reproduced the multiplicity fraction of star systems \citep{Bate2009a,Bate2012,KruKleMcK2012}. However, Figure \ref{mult} compares with the field multiplicity fraction, and it is not obvious why simulations should agree well with the field population while the cluster is gas dominated and not dynamically relaxed. If the initial cluster environment is initially dense, \citet{MocBat2010} demonstrated that the multiplicity remains nearly constant while the remaining gas is dispersed. \citet{ParReg2013} find that the shape of the mass ratio distribution is preserved during cluster relaxation because binary disruption is mass independent.  The distribution of multiples must also be reconciled with the stabilizing effect of magnetic fields \citep{ComHenHen2011,MyeMcKCunetal2013}, which appears to have a deleterious impact on small scale fragmentation and early multiplicity. 
In principle, the distribution of separations and mass ratios should provide strong constraints on simulations \citep{Bate2012} but are usually neglected. Future studies are required to determine how these properties depend on initial conditions and additional physics. More pointedly, future work must determine what such multiplicity statistics imply about the efficiency of star formation in dense gas and the origin of the IMF \citep{HolWalGooetal2013}.


\section{\textbf{SUMMARY}}\label{sum_sec}

Despite many lingering areas of debate surrounding the origin of the IMF, a few key areas enjoy consensus. The presence of supersonic turbulence plays an essential role in seeding structures that go on to spawn the stellar mass distribution.  Another fundamental area of physics is radiative transfer. A proper thermal treatment and heating from protostars tends to self-regulate the occurrence of additional fragmentation within cores and, when included in cluster simulations, reduces the numbers of sub-stellar objects and the number of stars per core, resulting in IMFs that agree well with the observed Galactic IMF. While the exact importance of magnetic effects remains poorly constrained, magnetization appears to have similar role in reducing fragmentation, while also creating outflows that entrain and eject core material, thus influencing the characteristic stellar mass.

In contrast, debate continues surrounding several other fundamental questions related to the IMF origin. For example, do the details and properties of the initial self-gravitating structures, ``cores'', have a unique and essential role in producing the IMF? The essence of the debate is one of nature versus nurture: do the initial dense cores play a greater role in determining stellar mass or does the environment and dynamical interactions dominate? This issue is obfuscated by the subtleties of defining cores observationally and connecting them to theoretical models. In recent years, simulations have shown that the combined influence of radiative and magnetic effects reduce fragmentation and hence dynamical interactions in the early stages of star formation, which makes stellar masses more similar to the masses of their parent cores; however, simulations have also shown that a core should not be viewed as a static object but as one that can gain mass on a dynamical timescale. 
Consequently, we conclude that both nature (cores) and nurture (dynamics) are involved to some extent in the origin of the IMF.  Future studies are needed to quantify this issue, especially with respect to connecting simulations more directly with observations.

An additional debate revolves around the meaningfulness of the similarity between the CMF and IMF, which seems to suggest a gas-to-star efficiency factor $\gtrsim$ 0.3. This efficiency likely results from some combination of multiplicity, mass loss from outflows, and dynamics; however, the interplay between these processes is highly nonlinear and exactly how each depends on core and star masses remains poorly understood. Future full-physics, parsec-size simulations of star formation with sub-AU resolution are necessary to disentangle these effects.

The fidelity of observations in both local and extreme regions has increased during the last decade, underscoring the apparent universality of the IMF across a wide range of environments to within observational uncertainty. This invariance suggests the IMF has minimal dependence on cluster size, cluster boundedness, and metallicity. Instances of proposed variation such as within the galactic center, giant ellipticals, and dwarf galaxies are tantalizing but tenuous. If the formation of stars is relatively insensitive to initial conditions and depends strongly on self-regulating fundamental physics as simulations and CMF models suggest then this would account for IMF invariance across a range of environments. Proposed theoretical solutions for the origin of IMF variance in extreme environments are still in their infancy. Future work is also needed to provide more robust constraints on observed low-mass stellar and core mass functions, natal conditions, and secondary characteristics such as multiplicity, which are necessary pieces of the theoretical puzzle of the IMF origin.

\textbf{ Acknowledgments.} The authors thank the referees, Ralph Pudritz and Ralf Klessen, for thoughtful comments that improved the review and Mark Krumholz, Andrew Myers, Philipp Girichidis, and Morten Andersen for sharing their data. Support for S.S.R.O. was provided by NASA through Hubble Fellowship grant HF-51311.01 awarded by the Space Telescope Science Institute, which is operated by the Association of  Universities for Research in Astronomy, Inc., for NASA, under contract NAS 5-26555. P.C.C. is supported by grant CL 463/2-1, part of the DFG priority program 1573 ``Physics of the Interstellar Medium'', and acknowledges financial support from the Deutsche Forschungsgemeinschaft (DFG) via SFB 881``The Milky Way System'' (sub-projects B1 and B2). P.H. has received funding from the European Research Council under the European Community's Seventh Framework Programme (FP7/2007-2013 Grant Agreement no. 306483).
N.B. acknowledges partial support from a Royal Society University Research Fellowship. E.M. acknowledges funding from the Agence Nationale de la Recherche contract ANR-2010-JCJC-0501-1 (DESC). APW thanks the UK STFC for support through grant PP/E000967/1.
\bigskip

\bibliographystyle{ppvi_lim1.bst}
\bibliography{bib_mbate,Off_IMF,Nate,PHOPKINS,PHENNEBELLE,Paul,estelle}

\end{document}